\documentclass[journal]{IEEEtran}
\usepackage{amsmath, amssymb, amsfonts}
\usepackage{mathtools}
\usepackage{braket}
\usepackage{graphicx}
\usepackage{threeparttable}
\usepackage{float, stfloats}
\usepackage{caption}
\usepackage{subcaption}
\usepackage{cuted}
\usepackage{color,colortbl}
\definecolor{lightgray}{gray}{0.9}
\definecolor{headerblue}{RGB}{64, 128, 255}
\renewcommand{\arraystretch}{1.3}  
\usepackage{array,tabularx,multirow}
\newcolumntype{P}[1]{>{\centering\arraybackslash}p{\#1}}
\usepackage{romannum}
\usepackage[T1]{fontenc}    
\usepackage[english]{babel} 
\usepackage{lettrine}
\usepackage{comment}
\usepackage{caption}
\usepackage{placeins}
\usepackage{balance}
\usepackage{array}
\usepackage{booktabs}
\usepackage{multirow}
\usepackage{tabularx}
\usepackage{booktabs}
\usepackage{caption}
\usepackage{ragged2e}
\usepackage{adjustbox}
\usepackage[dvipsnames]{xcolor}
\usepackage{pifont}
\usepackage{orcidlink}  
\usepackage{cuted}

\hypersetup{colorlinks=true,citecolor=black,linkcolor=black,urlcolor=black}
\usepackage[nameinlink,noabbrev]{cleveref}
\Crefname{figure}{Fig.}{Figs.}

\definecolor{IEEEblue}{RGB}{0,102,153}


\title{\textbf{\textcolor{IEEEblue}{Twin-Field Quantum Key Distribution: Protocols, Security, and Open Problems}}}

\author{
    \IEEEauthorblockN{Syed M. Arslan   \IEEEauthorrefmark{1}\orcidlink{0000-0002-5820-6576}}
    \IEEEauthorblockN{Syed Shahmir  \IEEEauthorrefmark{2}\orcidlink{0009-0001-1634-2713}}
    \IEEEauthorblockN{Noureldin Mohammad\IEEEauthorrefmark{1}}\orcidlink{0009-0001-4150-8690}
        \IEEEauthorblockN{Saif Al-Kuwari  \IEEEauthorrefmark{1}\orcidlink{0000-0002-4402-7710}} 
        \IEEEauthorblockN{Muataz Alhussein  \IEEEauthorrefmark{3}\orcidlink{0000-0002-1626-6511}}

    \IEEEauthorblockA{
        \IEEEauthorrefmark{1}Qatar Center for Quantum Computing, College of Science and Engineering, Hamad Bin Khalifa University, Doha, Qatar}\\
    \IEEEauthorblockA{
        \IEEEauthorrefmark{2}College of Science and Engineering, Hamad Bin Khalifa University, Doha, Qatar} \\
    \IEEEauthorblockA{
        \IEEEauthorrefmark{3}Department of Engineering, University of Cambridge, Cambridge, UK} 
}

\makeatletter
\renewcommand{\tableofcontents}{\section*{\contentsname\@mkboth{\contentsname}{\contentsname}}\@starttoc{toc}}
\makeatother

\begin{document}
\pagenumbering{arabic}
{\small
\maketitle}
\addtolength{\parskip}{-0.12em}

\begin{strip}
\noindent\textbf{\textcolor{IEEEblue}{Abstract} -}
Twin-Field Quantum Key Distribution (TF-QKD) has emerged as a potential protocol for long distance secure communication, overcoming the rate-distance limitations of conventional quantum key distribution without requiring trusted repeaters. By having two parties transmit phase encoded weak coherent pulses (WCP) to an untrusted central node, the TF-QKD exploits single-photon interference to achieve secret key rates scaling as square-root of channel length, enabling quantum-secured communication over unprecedented distances. This survey provides a comprehensive survey of TF-QKD, covering the original protocol, its fundamental principles, and key-rate derivation. We discuss major TF-QKD variants, including Phase-Matching QKD and Sending-or-Not-Sending QKD, with various improved versions. We compare their performance, implementation trade-offs, protocol-specific vulnerabilities, and countermeasures. The survey summarizes security proofs ranging from asymptotic decoy-state analyses to finite-key composable frameworks, experimental milestones, technological enablers, and practical deployment challenges. Finally, we outline open problems in the field and present a roadmap for integrating TF-QKD into scalable quantum networks, underscoring its central role in the future quantum internet.

\vspace{0.5em}
\noindent\textbf{\textcolor{IEEEblue}{Index Terms} -} Quantum Communication, Quantum Key Distribution, Twin-Field Quantum Key Distribution
\end{strip}


\begin{table*}[htbp]
\centering
\caption{Acronym Reference}
\label{tab:all_abbreviations_ultracompact_vertical}
\begin{tabular}{|p{1.2cm}|p{3.65cm}|p{1.2cm}|p{3.65cm}|p{1.2cm}|p{3.65cm}|}
\hline
\multicolumn{6}{|c|}{\textbf{Abbreviations Reference}} \\
\hline
\textbf{Acronyms } & \textbf{Full Form} & \textbf{Acronyms } & \textbf{Full Form}  & \textbf{Acronyms } & \textbf{Full Form} \\
\hline
AOPP & Active Odd-Parity Pairing & MITM & Man-in-the-Middle & APD & Avalanche Photodiode \\
NPS & No-Post-Selection & BB84 & Bennett-Brassard 1984 & OAM & Orbital Angular Momentum \\
BER & Bit Error Rate & OPLL & Optical Phase-Locked Loop & CKA & Conference Key Agreement \\
PLOB & Pirandola-Laurenza-Ottaviani-Banchi &
COW & Coherent One-Way & PM-QKD & Phase-Matching QKD \\
CV & Continuous Variable & PNS & Photon-Number-Splitting &
CV-QKD & Continuous Variable QKD \\
PTP & Precision Time Protocol &
DI-QKD & Device-Independent QKD & QBER & Quantum Bit Error Rate \\
DWDM & Dense WDM & QCKA & Quantum Conference Key Agreement &
E91 & Ekert 1991 \\ QKD & Quantum Key Distribution &
ECS & Entangled Coherent States & QND & Quantum Non-Demolition \\
ETSI & European Telecom Standards & QRNG & Quantum Random Number Generator &
 WDM & Wavelength Division Multiplexing  \\ RFI & Reference-Frame-Independent &
FPGA & Field-Programmable Gate Array & SDN & Software-Defined Networking \\
GLLP & Gottesman-Lo-Lütkenhaus-Preskill & SDQN & Software-Defined Quantum Networking &
HOM & Hong-Ou-Mandel \\ SNS & Sending-or-Not-Sending &
HSPS & Heralded Single-Photon Sources & SNSPD & Superconducting Nanowire SPD \\
SPE & Single-Photon Entanglement &
ITU-T & International Telecom Union & TF & Twin-Field \\
KMS & Key Management Systems & TF-QKD & Twin-Field QKD &
KPA & Known-Plain-text Attack \\
TLS & Transport Layer Security &
LEO & Low-Earth Orbit & VPN & Virtual Private Network \\
MAC & Message Authentication Code & WCP & Weak Coherent Pulse &
MDI & Measurement-Device-Independent \\ 
\hline
\end{tabular}
\end{table*}

\section{Introduction} \label{Intro}

Quantum key distribution (QKD) allows two distant parties to establish a shared secret key with security guaranteed by the fundamental laws of quantum mechanics. The foundational BB84 protocol, introduced by Bennett and Brassard in 1984 \cite{bennett_quantum_2014}, demonstrates this principle by exchanging single photons prepared in non-orthogonal polarization states over a lossy communication channel. The quantum no-cloning theorem ensures that any eavesdropping attempt necessarily disturbs these quantum states, creating detectable signatures of interception. This breakthrough sparked the development of numerous discrete-variable (DV) protocols, including Ekert's E91 protocol based on entanglement \cite{ekert_quantum_1991}, Bennett's two-state B92 \cite{bennett_quantum_1992}, and Bruss's six-state protocol \cite{bruss_optimal_1998}. In parallel, continuous-variable (CV) QKD emerged as an alternative approach, encoding information in the quadratures of coherent states and employing homodyne or heterodyne detection \cite{grosshans_continuous_2002}. Despite their information-theoretic security guarantees, practical QKD implementations face two fundamental challenges that have repeatedly limited their practical deployment. The first is the exponential decay of key generation rates with transmission distance due to optical fiber losses, creating a fundamental rate-distance trade-off bounded by the Pirandola-Laurenza-Ottaviani-Banchi (PLOB) limit \cite{pirandola_fundamental_2017}. The second stems from the gap between idealized theoretical models and realistic device implementations, which introduce security vulnerabilities through side channels and imperfections.

In addition to these deployment challenges, QKD faces other security ones. QKD's practical security concerns were first highlighted by photon-number-splitting (PNS) attacks, which exploit the multi-photon components inherent in WCP implementations of single-photon protocols \cite{avanesov_generalized_2025}. The decoy-state method, proposed by Hwang in 2003 \cite{hwang_quantum_2003} and rigorously developed by Lo \emph{et al.} in 2005 \cite{lo_decoy_2005}, provided an elegant solution by enabling precise characterization of single-photon contributions while maintaining the cost advantages of laser sources over true single-photon sources. Building on this foundation, measurement device-independent QKD (MDI-QKD) was introduced by Lo \emph{et al.} in 2012 \cite{lo_measurement-device-independent_2012} to address detector side vulnerabilities. By requiring both parties to send quantum states to an untrusted central measurement station, MDI-QKD eliminates all detector-side attacks while maintaining practical key rates. At the theoretical extreme, device independent QKD (DI-QKD) leverages Bell inequality violations to remove all assumptions about device implementations \cite{acin_device-independent_2007}, though at the cost of extremely low key rates and stringent experimental requirements for loophole-free Bell tests. While these advances significantly enhanced QKD security, they did not address the fundamental scaling limitation. Almost all protocols remained subject to linear scaling with channel transmittance ($R \sim \eta$, where $R$ is the secret-key generation rate and $\eta$ is the end-to-end channel transmittance), restricting secure communication to distances of a few hundred kilometers (without quantum repeaters). This limitation represented a critical barrier to the practical deployment of quantum-secured networks.

The Twin-Field QKD (TF-QKD), proposed by Lucamarini \emph{et al.} in 2018 \cite{lucamarini_overcoming_2018}, achieved a paradigm shift by fundamentally altering the rate-distance scaling relationship. Through a carefully designed protocol, where Alice and Bob each transmit WCP to an untrusted intermediate node, TF-QKD exploits single-photon interference to achieve key rate scaling as $R \sim \sqrt{\eta}$. This square root scaling represents a significant milestone that extends the reach of practical QKDs from hundreds to potentially thousands of kilometers. 

In this survey, we present a comprehensive examination of TF-QKD, starting with a detailed exposition of the original protocol and its theoretical foundations, then systematically reviewing major variants, analyzing security frameworks from asymptotic to finite-key regimes, chronicling experimental milestones, and outlining the path toward integrating TF-QKD into future quantum network infrastructures.


In the last couple of years, some reviews, tutorials, and survey articles have been published that cover the existing literature on TF-QKD.  Early practical surveys, such as the targeted review by Mao \emph{et al.} \cite{mao_recent_2021}, conducted a narrative-comparative review of protocols that exceeded the PLOB bound, with TF-QKD occupying a central position. This review provided the most complete account of TF-QKD and its variants, summarizing theoretical advances and fiber demonstrations $>$ 500 km. The discussion spans phase-matching, SNS, and decoy-state variants, supported by theoretical and experimental results. Similarly, Zhen \emph{et al.} \cite{yin_twin-field_2021}, developed a specialized perspective review devoted to the TF-QKD family for repeater-free intercity QKD, covering theoretical models, finite key analysis, and implementation pathways for multiple variants, including PM-QKD and no-phase post-selection schemes. Another comparative study by Yu \emph{et al.} \cite{yu_study_2021} offers a comparative review of several QKD protocols under realistic channel conditions, with TF-QKD as a key focus, detailing its original form, decoy state approach, and SNS variant while contrasting rate-distance performance with BB84 and MDI-QKD. They analyzed TF-QKD’s scaling advantage over other protocols, yet remained theoretical with asymptotic results, no real-world noise modeling, and no side-channel security assessment. 
Lauterbach \emph{et al.} \cite{lauterbach_study_2023}, adopt a narrative review format centered on satellite-based QKD, where TF-QKD is mentioned peripherally to contextualize performance benchmarks and decoy-state techniques. They focused on link budgets for satellite-based QKD architectures and noted the potential of TF-QKD for space links but did not offer experimental validation, satellite-specific phase stabilization modeling, or tailored finite key proofs. Around the same time, Dorozhynskyi \emph{et al.} \cite{dorozhynskyi_maximizing_2023} studied the integration of quantum cryptography into 5G networks, where TF-QKD is conceptually introduced as a long-distance candidate protocol without technical elaboration, lacking discussions on simulations, experimental metrics, and coexistence under mobile network conditions.  Similarly, Rajpurohit \emph{et al.} \cite{rajpurohit_advanced_2024}, presented a narrative review integrated with experimental system development, placing TF-QKD and MDI-QKD at the core of their discussion, and covering multiple variants, including discrete-phase-randomized, phase-matching, and post-selection strategies, alongside implementation challenges and countermeasures. The review lacked large-scale or long-distance field trials, finite-key security analysis, and benchmarking of phase drift compensation methods under realistic noise. In the urban networking context, Granados \emph{et al.} \cite{granados_quantum_2025}, identified TF-QKD’s role in extending metropolitan backbones but lacked real deployment trials, coexistence studies with classical WDM traffic, and multi-user finite-key security proofs. Collectively, most of these studies range from broad application-oriented overviews, where TF-QKD appears only as a contextual note, to highly focused studies offering detailed comparative analysis of its operational principles, variants, and experimental realizations.

While these surveys collectively provide valuable perspectives on QKD and, to varying degrees, TF-QKD, several gaps remain unaddressed. First, no existing work offers a unified, systematic comparison of all major TF-QKD variants, i.e., original, phase-matching, SNS, discrete-phase-randomized, and no-phase-postselection under a consistent performance and security framework, particularly in the finite-key regime. Second, the integration of TF-QKD into broader quantum network and multiparty communication scenarios remains largely unexplored, with most prior reviews focusing on point-to-point or specialized application contexts (e.g., satellite, 5G, or metropolitan networks) without cross-domain synthesis. Third, experimental progress is often presented in isolation, lacking a consolidated timeline and direct correlation with theoretical advances, which is essential to identify maturity stages and levels of technological readiness. Finally, coherent attack models and practical security considerations specific to TF-QKD, such as phase reference sharing, mode mismatch, and channel asymmetry, are either treated superficially or omitted. This survey addresses these gaps by providing a comprehensive, variant-inclusive analysis of TF-QKD, mapping theoretical developments to experimental milestones, and situating the protocol family within emerging quantum network architectures while explicitly evaluating its resilience to advanced adversarial strategies. Table \ref{tab:tfqkd_surveys_matrix} provides a comprehensive comparison between existing reviews (discussed above), revealing clear theoretical gaps that are filled by this survey.

\begin{table*}[!t]
\centering
\caption{Comparison of the survey on QKD with emphasis on twin-field QKD (TF-QKD).}
\label{tab:tfqkd_surveys_matrix}
\renewcommand{\arraystretch}{1.3}
\begin{tabularx}{\linewidth}{|p{1.8cm}|X|X|X|X|p{1.8cm}|X|p{1.8cm}|}
\hline 
\multirow{2}{*}{\textbf{Reference}} &
\multicolumn{3}{c|}{\textbf{TF-QKD Variants}} & 
\multicolumn{4}{c|}{\textbf{Coverage Areas}} \\ 
\cline{2-8}
&  \textbf{Original TF} & \textbf{PM-QKD} & \textbf{SNS-TF} & \textbf{Security} & \textbf{Experiments} & \textbf{Networks} & \textbf{Implementation} \\
\hline
\cite{mao_recent_2021} (2021)  & \checkmark & \checkmark & \checkmark & \checkmark & \checkmark & \texttimes & \checkmark \\
\hline
\cite{yu_study_2021} (2021) & \checkmark & \checkmark & \texttimes & \checkmark & \checkmark & \texttimes & \texttimes \\
\hline
\cite{lauterbach_study_2023} (2023) & \checkmark & \texttimes & \texttimes & \texttimes & \checkmark & \checkmark & \texttimes \\
\hline
\cite{dorozhynskyi_maximizing_2023} (2023) & \checkmark & \texttimes & \texttimes & \texttimes & \texttimes & \checkmark & \texttimes \\
\hline
\cite{rajpurohit_advanced_2024} (2024) & \checkmark & \texttimes & \texttimes & \texttimes & \checkmark & \texttimes & \checkmark \\
\hline
\cite{granados_quantum_2025} (2025) &\texttimes & \texttimes & \checkmark & \texttimes & \checkmark & \checkmark & \checkmark \\
\hline
This Survey & \checkmark & \checkmark & \checkmark & \checkmark & \checkmark & \checkmark & \checkmark \\
\hline
\end{tabularx}
\end{table*}

\subsection{Scope}
\begin{figure}[H]
    \centering
    \includegraphics[width=0.9\linewidth]{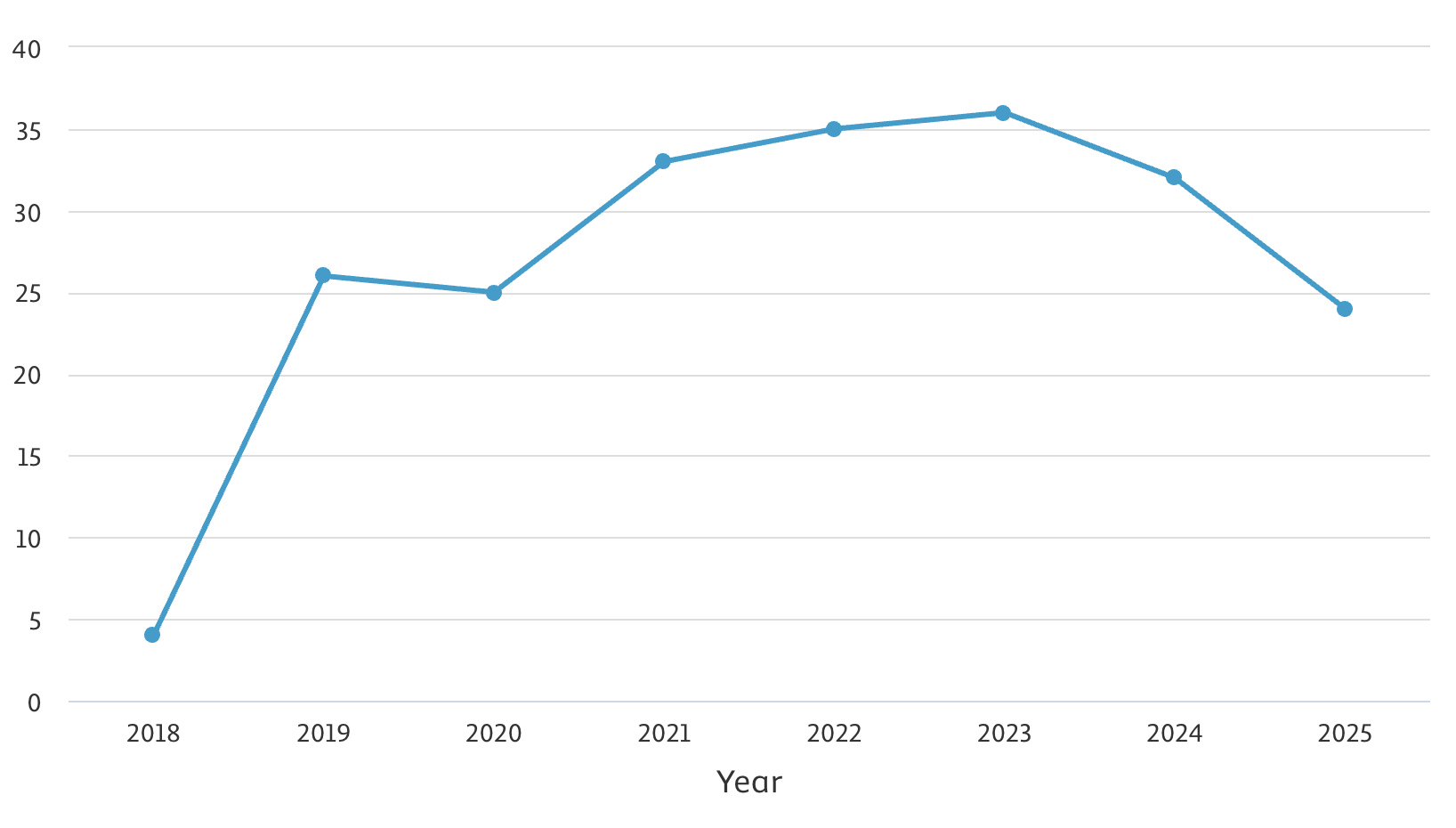}
    \caption{Number of Publications per year.}
    \label{fig:pubperyear}
\end{figure}

TF-QKD addresses challenges beyond extending distance: it targets network scalability, operational complexity, and the commercial viability of quantum-secured communications. By surpassing the linear rate-loss scaling (the PLOB bound) for repeaterless channels, TF-QKD achieves the square-root scaling \(R \propto \sqrt{\eta}\), where \(\eta\) is the end-to-end channel transmittance, and has enabled practical secure links beyond 500 km in fiber \cite{chen_sending-or-not-sending_2020}. In fact, TF-QKD initiated new avenues to explore inter-city and potentially continental-scale quantum networks without relying on full quantum repeaters. 
The selected literature is classified into four categories: theoretical protocols and variants (\(48\%\)), security analysis and proofs (\(20\%\)), experimental demonstrations and validation (\(21\%\)), and practical implementation considerations (\(11\%\)). This taxonomy traces the field’s evolution from early theory to deployment readiness, with a recent shift toward finite-key security and real-world validation in Fig. \ref{fig:pubperyear}.

\begin{figure}[H]
    \centering
    \includegraphics[width=0.9\linewidth]{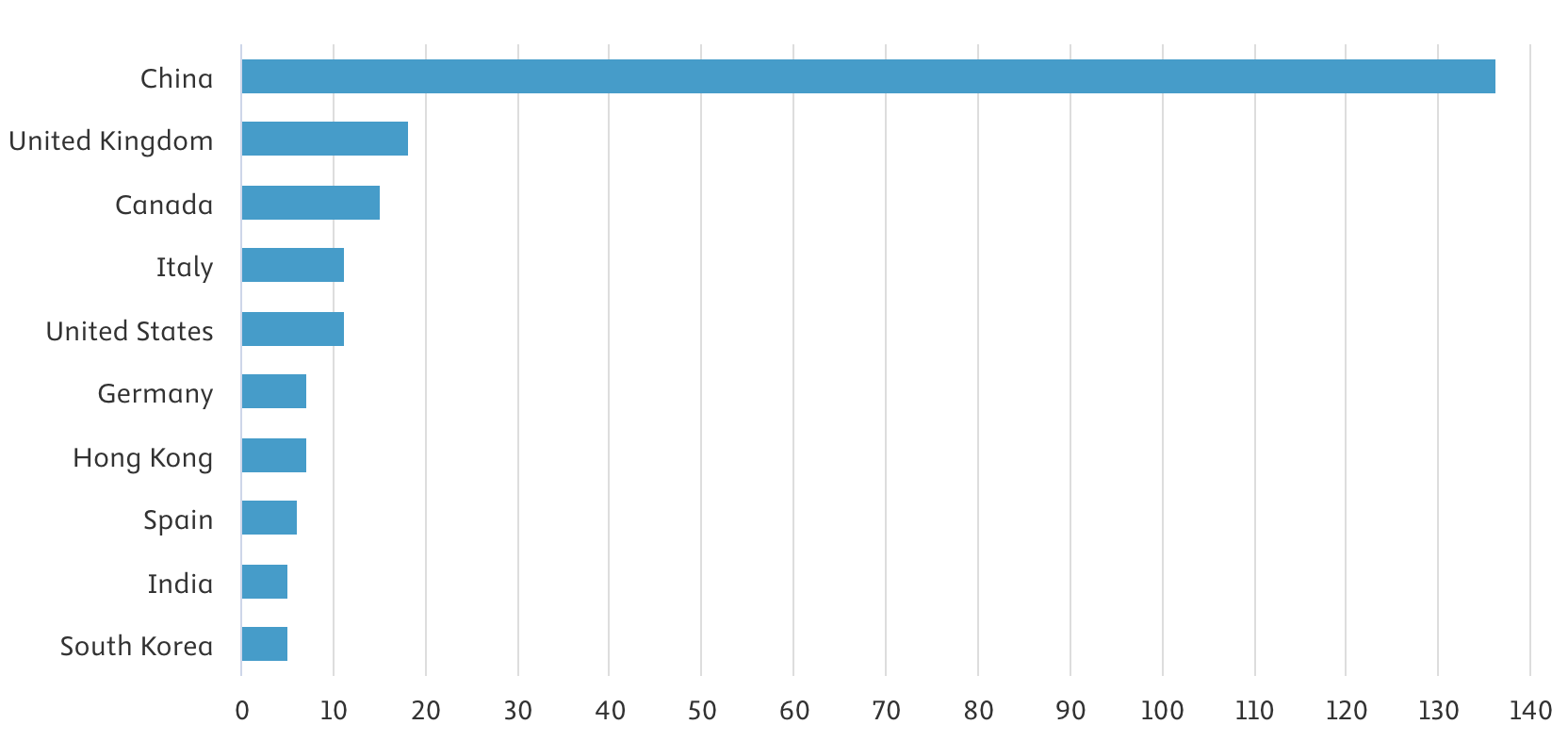}
    \caption{Geographical coverage of as per total number of publications.}
    \label{geo}
\end{figure}

The temporal scope spans foundational QKD work from the 1984 BB84 protocol through the 2017 PLOB rate-loss bound, providing the context needed to assess TF-QKD’s impact. Geographical coverage in Fig. \ref{geo} reflects global contributions and highlights leading groups and institutions that have driven progress to date,

\begin{figure}[H]
    \centering
    \includegraphics[width=1\linewidth]{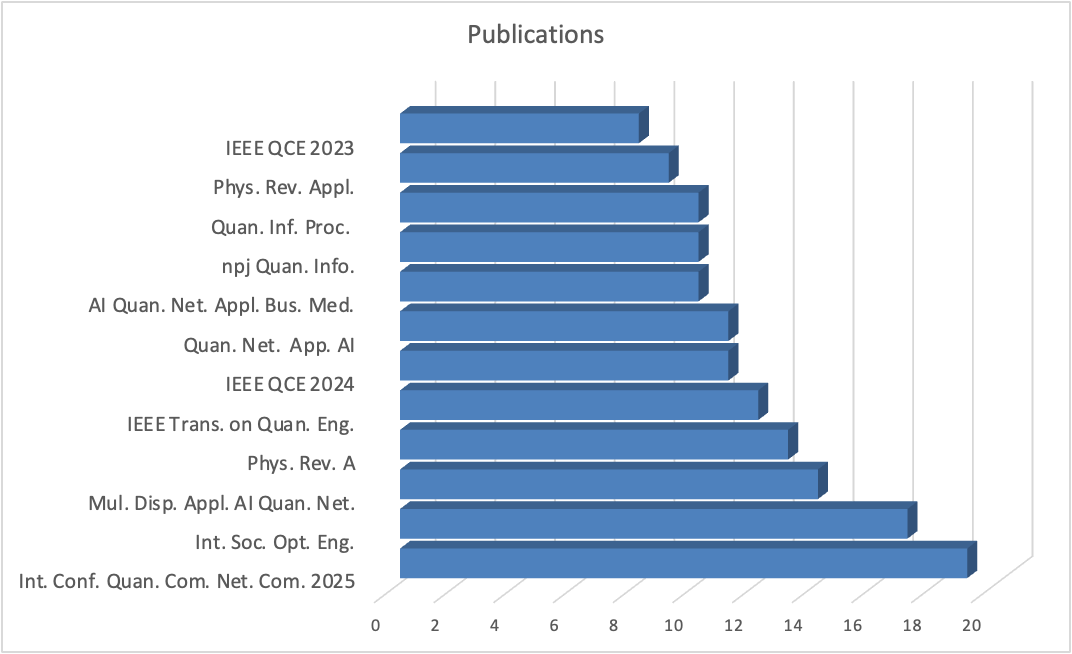}
    \caption{Distribution of TF-QKD publications across the top-12 venues in our corpus.}
    \label{fig:Piechart}
\end{figure}

Furthermore, Fig. \ref{fig:Piechart} shows the distribution of TF-QKD papers across the top 12 publication venues. Percentages denote the share of each venue within the data set; abbreviations are used for brevity.

Furthermore, this work integrates theoretical advances, security models, and experimental developments. We address the central challenge of twin-field quantum key distribution (TF-QKD) within the evolving landscape of quantum communication networks. The main target is the comprehensive understanding of how TF-QKD and its variants can be optimized, implemented, and secured in both current and next-generation quantum networks. Followed by a detailed discussion on the operational, security, and integration aspects of TF-QKD.

\subsection{Contributions}

This survey makes four contributions to the TF-QKD literature:
\begin{enumerate}
    \item We provide a comprehensive classification of TF-QKD variants, aligned with their performance metrics, security characteristics, and implementation complexity assessments, which supports informed protocol selection.
    \item We connect asymptotic theory with finite-key practical requirements via a unified security analysis, clarifying measurement-device-independent principles while addressing protocol-specific vulnerabilities.
    \item We correlate theoretical advances with experimental milestones to identify the factors that enabled the rapid progress of TF-QKD and the technological developments needed next.
    \item We propose practical deployment frameworks covering hardware, network architectures, and integration challenges grounded in field-trial experience to guide the transition from research demonstrations to commercial quantum-secured networks.
\end{enumerate}

\subsection{Organization}

The rest of this survey is organized as follows: Section \ref{CovQKD} establishes the theoretical foundations of QKD. Section \ref{QKD_Secure} outlines the security frameworks relevant to its advantages. In Section \ref{QKD_Challenge}, we present the fundamental challenges faced by various QKD schemes and the motivation for adopting the TF paradigms. Furthermore, Section \ref{TF-QKD} explains in detail the TF-QKD schemes, formalizes the security assumptions, motivations for the main variants, and provides the original protocol description with a step-by-step key-rate derivation to clarify both mechanisms and analytical techniques. Section \ref{TF-Var} surveys the landscape of TF-QKD variants, grouped into three families: Phase-Matching QKD (PM-QKD), Sending-or-Not-Sending (SNS) protocols, and hybrid TF-QKD variants, along with TF-QKD with additional performance/security improvements. For each, we summarize the core ideas, performance characteristics, and implementation trade-offs, yielding a practical taxonomy. Section \ref{TF:security_proofs} provides a security analysis in these families, contrasting asymptotic and finite-key regimes, detailing protocol-specific vulnerabilities, and highlighting countermeasures. Furthermore, Section  \ref{TF_experiment} \& \ref{TF:Implement} consolidate experimental progress, technological developments, and implementation along with practical deployment considerations with current infrastructure. Section \ref{TF-Networks} extends to the utility and application of TF-QKD in quantum network and multiparty cases, followed by Section \ref{open_problems}, where open problems and future directions are discussed. Finally, concluding in Section \ref{conclusion}.


\section{Conventional QKD} \label{CovQKD}

Quantum key distribution (QKD) was invented in 1984 by Bennett \& Brassard with their seminar work that led to the BB84 protocol \cite{bennett_quantum_2014}. In BB84, Alice encodes classical bits into non‐orthogonal polarization states of single photons and Bob measures them in randomly chosen bases.  Its security hinges on the no-cloning theorem, where any eavesdropper, Eve, unavoidably disturbs those states, introducing detectable errors during sifting and reconciliation \cite{bennett_quantum_2014}. Building on it, Ekert’s E91 entanglement-based protocol replaced direct preparation with shared Bell pairs, using Bell inequality violations as a security certificate and removing trust in state preparation \cite{ekert_quantum_1991}. Shortly thereafter, Bennett’s B92 protocol, introduced in 1992, takes BB84 prepare-and-measure concept and strips it down to its bare essentials and includes entanglement. In the B92, Alice uses only two non-orthogonal states (for example, horizontal and +45$^o$) and  Bob randomly chooses one of two measurement bases that are each orthogonal to one of Alice’s sent states; a click in Bob’s detector therefore unambiguously informs him of Alice's transmitted message. Any eavesdropper’s attempt to distinguish them inevitably introduces detectable errors, but the protocol’s extreme simplicity comes at the cost of a lower tolerable error rate and reduced key rate compared with BB84 \cite{bennett_quantum_1992}. In 1998,  Bruss \cite{bruss_optimal_1998}, proposed the six-state protocol, in which Alice prepares her single photons in one of six equally spaced polarization states drawn from the three mutually unbiased bases (horizontal/vertical, diagonal/antidiagonal, and right-/left-circular) and Bob likewise measures in one of these three bases at random.  By tripling the number of possible states and measurement choices, the six-state scheme tightens the allowable eavesdropper disturbance across three bases instead of two. Hence, the key remains secure even at higher noise levels. However, the additional bases mean Alice and Bob discard a larger fraction of their raw data (whenever Bob’s basis does not match Alice’s), which can reduce the raw key rate under low-noise conditions. The major issue in the original BB84 was in its experimental implementation, as it was susceptible to the photon-number-splitting (PNS) attack, which exploits the fact that WCP sometimes contains more than one photon.. In a PNS attack, Eve selectively captures extra photons from multi-photon pulses, learning part of the key without inducing any error in the single-photon events Bob registers, thus undermining the security of BB84 \cite{lutkenhaus_quantum_2002}. For this, the Decoy state QKD \cite{lo_decoy_2005} was invented to thwart PNS attack. 
The decoy-state method remedies this by having Alice randomly vary the intensity of her pulses among several values (signal, decoy, and sometimes vacuum levels) while keeping their encoding procedure identical. Since Eve cannot distinguish signal pulses from decoys in real time, any attempt to preferentially split multi-photon signals will alter the detection statistics (e.g., the gain and error rate) of the different intensity classes in a detectable way. By comparing the measured detection outputs (observed yields) i.e. Bob gets a detection event for a given class of sent pulses by Alice, and quantum bit error rates of signal and decoy states, Alice and Bob can tightly bound the fraction of single‐photon detections and thus accurately estimate Eve’s information. This restoration of a reliable single‐photon channel allows secure key rates to scale linearly with channel transmittance $\eta$ effectively, regaining the full distance potential of BB84 and its variants while using practical laser sources \cite{wang_beating_2019, zhao_experimental_2006}. 

In parallel, continuous-variable QKD (CV-QKD) encoded keys in quadratures of weak coherent state and its experimental implementation, i.e. phase or amplitude. The Gaussian-modulated CV-QKD protocol was first demonstrated by Grosshans and Grangier in 2002 \cite{grosshans_continuous_2002}, where Alice draws two real numbers, i.e. $\alpha = x + \iota p$, and sends them over the quantum channel to Bob, who performs homodyne or heterodyne detection to measure one or both quadratures of the incoming state. The raw correlated data are then sifted, error corrected, and privacy amplified to produce a secure key. In 2013, Leverrier \emph{et al.} \cite{leverrier_security_2013} provided security proofs of CV-QKD against collective and coherent attacks. The experimental advancement of QKD progressed rapidly from lab proof-of-principle to real-world demonstration, and in the 1990s, Buttler \emph{et al.} tested over a 205m indoor optical path at Los Alamos National Laboratory \cite{buttler_free-space_1998}, which was later extended in Geneva to 23 km deployment over telecom fibers with fringe visibility of 0.9984 \cite{muller_plug_1997}. 
Followed by the first experimental implementation of decoy-state protocols in 2006 \cite{zhao_experimental_2006}, enabling secure long-distance QKD. Continuous improvements in detector efficiency, digital signal processing, and error-correction codes pushed the fiber transmission distances beyond 80 km \cite{takesue_quantum_2007, lydersen_superlinear_2011, peev_secoqc_2009}. More recent advances have explored measurement-device-independent QKD (MDI-QKD), extended up to 200km, improving the security by removing trust from detectors \cite{tang_measurement-device-independent_2014}. MDI-QKD showed feasibility over metropolitan distances with coherent states and Gaussian modulation \cite{jouguet_analysis_2012, sun_analyzing_2025, fletcher_overview_2025}. Similar schemes for satellite links and integration on photonic chips have also been explored, leveraging compatibility with standard telecom hardware, high clock rates, and ease of multiplexing to deliver secure keys over global networks \cite{liao_satellite-relayed_2018, yin_entanglement-based_2020}.

\begin{figure}
    \centering
    \includegraphics[width=1.01\linewidth]{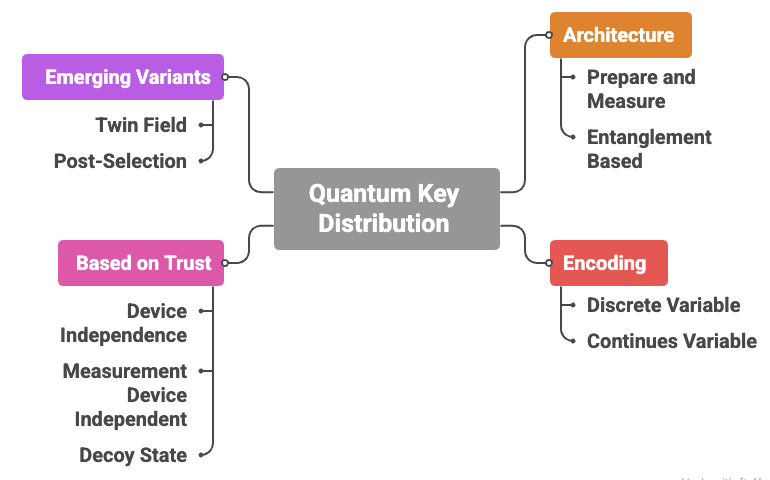}
    \caption{Hierarchical taxonomy of Quantum Key Distribution (QKD) protocols, organized into three overarching categories.}
\end{figure}

\begin{figure}
    \centering
    \includegraphics[width=1\linewidth]{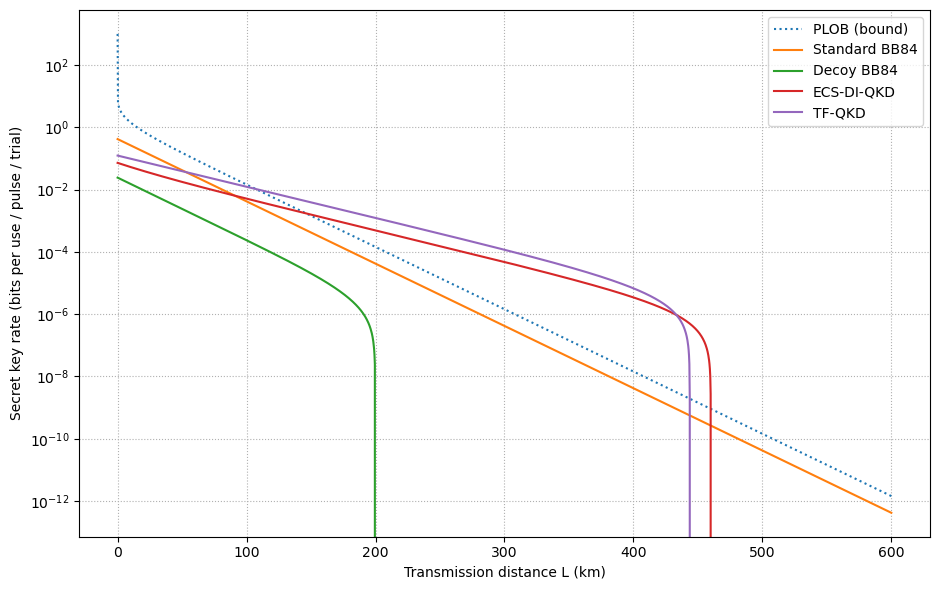}
    \caption{Comparison of secret key rate versus transmission distance for various QKD schemes. The plot shows the performance of standard BB84, Decoy-BB84, ECS-DI-QKD, and TF-QKD under different assumptions, with a channel loss coefficient of $\alpha$ = 0.2 dB/km and the log of secret key rate vs the transmission distance. The curves illustrate the trade-off between key rate and maximum achievable distance for each scheme and baseline against the PLOB bound.}
    \label{PLOB}
\end{figure}

In addition to classification by protocol type, QKD implementations can also be broadly divided into active and passive schemes, depending on how quantum states are prepared and measurement bases are selected \cite{ma_alternative_2012, makarov_effects_2006, teng_twin-field_2020}. 
\begin{itemize}
    \item In active schemes, high-speed modulators driven by random number generators dynamically choose the basis or encoding of each signal, offering precise control and adaptability but potentially introducing modulator-related side channels. 
    \item In passive schemes, the basis choice is determined inherently by quantum randomness, e.g., through beam splitters that route photons along different optical paths, reducing electronic complexity and certain side-channel risks, often at the cost of additional optical loss and lower raw detection efficiency. 
\end{itemize}
Both approaches are compatible with multiple QKD protocols, and the choice typically reflects trade-offs between performance, hardware complexity, and implementation security \cite{jofre_active_2011, wang_fully_2023}. 
The evolution of QKD has produced a diverse landscape of protocols, each addressing specific security challenges and practical constraints. Table \ref{tab:unified_qkd_comparison1} provides a comprehensive comparison of the main conventional QKD protocols, highlighting their key characteristics, security model, and rate-distance limitations. Conventional protocols such as BB84, E91, and Decoy-state QKD are well-established and typically operate over 100-300 km, with linear key rate scaling ($R \sim \eta$). Advanced QKD protocols introduce stronger security models or novel paradigms, but may incur higher complexity or distance limitations. In recent years, new approaches have aimed to enhance performance beyond traditional rate-loss trade-offs, achieving improved key rate scaling and extended secure transmission distances. These methods leverage coherent-state encoding, phase correlations, and optimized detection schemes to overcome the fundamental rate-distance barrier predicted by the PLOB bound, as illustrated in Fig. \ref{PLOB}. As summarized in Table \ref{tab:unified_qkd_comparison}, several modern coherent-state-based QKD variants such as phase-matching QKD (PM-QKD), sending-or-not-sending (SNS) protocols, and no post-selection schemes use WCPs and single-photon interference at an intermediate node to extend secure communication well beyond 500 km. With moderate to high implementation complexity, these next-generation QKD approaches are well suited to support future intercity quantum networks. We discuss these protocols and their performance characteristics in detail in Section \ref{TF-QKD} \& \ref{TF-Var}. 

\begin{table*}[htbp]
\centering
\caption{Comprehensive comparison of quantum key distribution protocols}
\label{tab:unified_qkd_comparison1}
\footnotesize
\begin{tabular}{|p{2.1cm}|p{1.3cm}|p{2.1cm}|p{1.8cm}|p{1.6cm}|p{2.2cm}|p{3.4cm}|}
\hline
\textbf{Protocol} & \textbf{Key Rate Scaling} & \textbf{Security Model} & \textbf{Source Type} & \textbf{Typical Distance} & \textbf{Implementation Complexity} & \textbf{Key Characteristics} \\
\hline
\multicolumn{7}{|c|}{\textbf{Discrete-Variable QKD Protocols}} \\
\hline
BB84 & $R \sim \eta$ & Prepare-measure & Single photon/WCP & 100--200 km & Moderate & Simple, well-established; vulnerable to detector attacks \\
\hline
E91 & $R \sim \eta$ & Entanglement-based & Entangled pairs & 100--200 km & High & Bell inequality security; requires high-quality entanglement \\
\hline
B92 & $R \sim \eta$ & Prepare-measure & Single photon/WCP & 100--200 km & Low & Two-state simplicity; lower information efficiency \\
\hline
Six-state & $R \sim \eta$ & Prepare-measure & Single photon/WCP & 100--200 km & Moderate & Enhanced eavesdropping detection; higher QBER tolerance \\
\hline
Decoy-state QKD & $R \sim \eta$ & Prepare-measure & WCP with decoys & 200--300 km & Moderate & Practical laser sources; PNS attack resistance \\
\hline
\multicolumn{7}{|c|}{\textbf{Continues-Variable QKD Protocols}} \\
\hline
Gaussian-Modulated & $R \sim \eta$ & Continuous variables & Coherent states & 80-150 km & Moderate & Uses homodyne/heterodyne detection; telecom-compatible; mature security proofs \\ \hline
Discrete-Modulated & $R \sim \eta$ & Continuous variables & Coherent states & 80-150 km & Moderate & Lower reconciliation cost; simpler modulation hardware \\
\hline
\end{tabular}
\end{table*}

\section{QKD Security} \label{QKD_Secure}

The transition from theoretical protocols to practical implementations revealed significant security gaps. QKD is often presented as \emph{unconditionally secure}, but a detailed critical analysis reveals that these security claims are sometimes overstated and several key vulnerabilities remain unsolved \cite{yuen_security_2016}. Early QKD systems relied on ideal assumptions where the actual security provided by a QKD system is not the theoretical unconditional security of the laws of physics, but rather the more limited security that can be achieved by hardware and engineering designs \cite{metger_security_2023}. Early proofs relied on mutual information and later on the trace distance to quantify secrecy. The evolution of QKD security research has followed a chronological pattern of vulnerability discovery followed by countermeasure development, creating an ongoing cycle that continues to refine practical quantum cryptographic systems.

The foundational security challenges rose in the 1990s, as it became clear that QKD (like classical systems) is vulnerable to man-in-the-middle (MITM) attacks when operated without authentication, since no known principle of quantum mechanics can distinguish a legitimate party from an adversary. This vulnerability affects virtually all known QKD schemes without authentication \cite{fei_quantum_2018}. Consequently, an initial shared secret or an authentication mechanism is required, as Alice and Bob have no means of verifying each other’s identities otherwise. A standard solution is to combine QKD with unconditionally secure authentication schemes, such as the Carter-Wegman method, enabling exponential expansion of an initial short key. However, in the case of Carter-Wegman, achieving unconditional security still requires the use of strong universal families of hash functions, which ensure that the probability of an adversary successfully forging a valid message-tag pair is strictly bounded and independent of computational assumptions.
In the 2000s, research into quantum information locking demonstrated that partial knowledge of the key generated through QKD can reveal substantial information in a known-plain-text attack (KPA). This finding directly exposed weaknesses in the assumption that a small trace distance value automatically ensures strong subsequence security \cite{yuen_security_2016, divincenzo_locking_2004}. In 2003, research by NIST showed that QKD is vulnerable to a classic man-in-the-middle attack using entangled pairs created by Eve \cite{kuhn_vulnerabilities_2003}. These early attacks could be applied to any protocol that relies on the manipulation and return of entangled qubits to create a shared key. This led to a substantial revision and research on the fundamentals of QKD protocols. A major theoretical milestone occurred in 2014 with the Pirandola-Laurenza-Ottaviani-Banchi (PLOB) bound, which set a fundamental limit on the secret key capacity of lossy channels, reframing expectations for the achievable rates and distances in practical QKD systems \cite{pirandola_fundamental_2017, pirandola_advances_2020}. In the 2010s, a deeper analysis revealed that the common interpretation of trace distance to quantify secrecy as a failure probability was mathematically incorrect, and that even extremely small values, such as $10^{-20}$, can be inadequate in high-volume deployments due to the accumulation of probability on many keys \cite{renner_security_2008}. These critiques also highlighted incomplete modeling of error correction and privacy amplification, which can leak side information to an eavesdropper, especially as the literature lacked rigorous bounds on bit error rates (BER) \cite{tsurumaru_leftover_2020}. Today, the consensus is shifting toward using operational, probability-based measures of security that directly bound adversary success in guessing full or partial keys, incorporating realistic physical limits such as the PLOB bound, and ensuring strong, composable authentication to defend against MITM attacks. Addressing these issues is essential to bridge the gap between the theoretical promise of QKD and its proven, deployable security in real-world systems. This foundational understanding established the importance of considering practical implementation vulnerabilities beyond theoretical security proofs. 

In the following, we discuss in detail popular attacks against QKD systems, highlighting potential countermeasures.

\subsubsection{Man-in-the-Middle (MITM) Attacks}
A QKD system, while inherently secure against many eavesdropping strategies, is still vulnerable to a classical MITM attack if authentication is not implemented. In such a scenario, an adversary (commonly referred to as Eve) can intercept the quantum channel and impersonate each legitimate party to the other. Eve could, for example, receive quantum states sent by Alice, measure them, and then prepare and transmit new quantum states to Bob. Simultaneously, she would communicate with Alice while pretending to be Bob. Since the quantum channel alone does not verify the identity of its participants, both Alice and Bob could be tricked into believing they are exchanging qubits with each other when, in fact, all communication is routed through and manipulated by Eve. This breaks the integrity of the protocol before the benefits of quantum security can even take effect \cite{fei_quantum_2018, amellal_quantum_2023}.

To counteract this vulnerability, authentication mechanisms must be applied to the classical communication channel that accompanies QKD. One robust approach is to use information-theoretically secure message authentication codes (MACs) based on pre-shared secret keys. Such MACs provide unconditional security against forgery, ensuring that any attempt by an adversary to alter or fabricate messages is detectable, regardless of computational power. The initial authentication key can be relatively short and reused to authenticate subsequent QKD sessions, with fresh keys generated by QKD itself. Alternatively, post-quantum secure authentication methods, such as those based on lattice-based, code-based, hash-based cryptography or the FALCON algorithm can be employed to resist future quantum computing threats \cite{alhazmi_mitigating_2023, bibak_quantum_2021}. These techniques ensure that both quantum and classical exchanges occur strictly between legitimate parties, effectively nullifying the risk of MITM attacks and preserving the overall security promise of QKD.

\subsubsection{Photon Number Splitting (PNS) Attacks}
The PNS attack emerged as one of the most significant practical vulnerabilities in QKD implementations during the late 1990s and early 2000s. The PNS Attack was first described by Brassard, Lütkenhaus, Mor and Sanders in a 1999 paper \cite{brassard_limitations_2000}, exploiting the fundamental limitation that practical QKD systems cannot generate perfect single-photon sources \cite{lutkenhaus_security_2000}. In practice, many implementations use laser pulses attenuated to a very low level to send the quantum states, with these laser pulses containing a very small number of photons distributed according to a Poisson distribution, meaning most pulses contain no photons, some contain one photon, and a few contain two or more photons. The attack mechanism involves Eve intercepting multi-photon pulses, splitting off extra photons while transmitting the remaining single photon to Bob. Eve stores these extra photons in a quantum memory until Bob detects the remaining single photon and Alice reveals the encoding basis, allowing Eve to measure her photons in the correct basis and obtain information on the key without introducing detectable errors \cite{brassard_limitations_2000, lutkenhaus_security_2000}. This vulnerability posed a critical threat to the security assumptions underlying practical BB84 implementations using WCP rather than ideal single-photon sources.

The development of countermeasures against PNS attacks led to several breakthrough approaches. The Gottesman-Lo-Lütkenhaus-Preskill (GLLP) framework addressed these imperfections by deriving secure key rate bounds under limited PNS scenarios, providing a theoretical foundation for security analysis with imperfect sources \cite{hwang_quantum_2003, gottesman_security_2004}. More significantly, the decoy-state protocol emerged as a practical solution, with the decoy-state protocol improving performance by refining single-photon yield estimates and enabling the detection of photon-number-splitting eavesdropping attempts in practical systems \cite{hwang_quantum_2003, wang_beating_2005}. The decoy-state method thwarts PNS attacks by varying the mean photon number between pulses, enabling detection of anomalous photon statistics caused by an attacker. This technique restores unconditional security for WCP-based QKD within the appropriate channel loss constraints.

\subsubsection{Trojan Horse Attacks}
Trojan horse attacks are an active side-channel threat in QKD systems, where an eavesdropper (Eve) deliberately injects bright optical pulses into the quantum channel to probe the internal state of the target device. Instead of attacking the quantum states themselves, which are protected by the no-cloning theorem, these attacks exploit device imperfections to extract classical side-channel information, such as basis choices, phase modulation values, or polarization settings \cite{gisin_trojan-horse_2006}. When Eve’s injected light enters the device, part of it can be reflected or back-scattered from optical components, including fiber connectors, beam splitters, modulators, or mirrors. This returning light can carry subtle imprints of the device’s internal configuration, which Eve can analyze to infer secret settings \cite{vakhitov_large_2001}. Variants of this strategy include reflection-based probing using Fresnel or Rayleigh back-scattering, multi-wavelength injection to bypass narrow-band spectral filters, time-domain probing to interrogate modulators during specific switching intervals, and exploitation of nonlinear optical effects induced by high-intensity illumination \cite{lydersen_hacking_2010}. These techniques are especially dangerous because they target the physical implementation of QKD rather than its abstract protocol, meaning that conventional security proofs may not apply unless such leakage channels are explicitly modeled.

Mitigating Trojan horse attacks requires a combination of hardware countermeasures, monitoring systems, and protocol-level defenses. Broadband optical isolators should be installed at the channel input to block incoming light over a wide spectral range, preventing multi-wavelength bypass attacks \cite{makarov_effects_2006}. High-speed photo-detectors can be used to continuously monitor incoming power levels, with alarms triggered when the detected power exceeds the expected single-photon or faint-pulse regime. Multi-wavelength filtering, such as with dense wavelength division multiplexing (DWDM) filters or broadband rejection filters, can block unauthorized spectral components. Additionally, randomizing internal modulation parameters, even introducing dummy settings, can make any leaked information unusable to Eve \cite{lucamarini_efficient_2013}. Finally, for true end-to-end security, isolation performance, power monitoring specifications, and allowable leakage thresholds must be incorporated into the formal security proofs of the QKD system. By embedding these protections both in hardware and in the theoretical model, the risk from Trojan horse attacks can be significantly reduced \cite{peng_security_2024}.

\subsubsection{Detector Control and Blinding Attacks}
Avalanche photodiode (APD) detectors, which are commonly used in QKD systems, can be vulnerable to a class of attacks known as detector control and blinding attacks. In such attacks, an eavesdropper (Eve) can shine strong continuous-wave or pulsed light onto the APDs, forcing them out of their intended Geiger mode of single-photon sensitivity and into a linear, classical response regime. Once in this regime, detectors no longer register genuine quantum signals, but instead respond predictably to classical light intensities \cite{makarov_effects_2006, navas-merlo_detector_2021}. This enables Eve to send carefully tailored “faked states” that determine exactly which detector will register a click, effectively allowing her to control the measurement outcomes without introducing detectable anomalies in the quantum bit error rate (QBER). A variation of this method exploits the gating windows of the detectors in gated-mode QKD systems. By timing the blinding pulses to coincide with specific gate intervals, Eve can manipulate the detection events in her favor while avoiding suspicion \cite{lydersen_hacking_2010}.

To defend against these vulnerabilities, several countermeasures have been proposed and implemented. One approach is to randomize detector parameters, such as bias voltages or detection efficiencies, making it difficult for Eve to predict the detector response and calibrate her attack. Another strategy involves incorporating watchdog photo-diodes that monitor incoming optical power, enabling the system to detect and respond to abnormal high-power illumination indicative of a blinding attempt \cite{lydersen_avoiding_2010}. However, the most robust solution is to adopt MDI-QKD protocols, which eliminate the need for trusted detectors entirely by shifting the measurement process to an untrusted relay. This design inherently removes all detector side-channels, ensuring security even in the presence of compromised or malicious measurement devices.

\subsubsection{Time-Shift and Efficiency Mismatch Attacks}
Time-shift and efficiency mismatch attacks exploit imperfections in the single-photon detectors used in QKD systems. In practice, detectors often have slightly different efficiencies, and these efficiencies can vary depending on the precise arrival time of the photon. An eavesdropper, Eve, can take advantage of this by deliberately introducing a time shift in the incoming photons, thereby influencing which detector is more likely to register a click. By correlating her timing manipulation with the detector that fires, she can gain partial information about the key without introducing a noticeable level of QBER that would alert legitimate parties \cite{xu_secure_2020}. 

To defend against such attacks, several countermeasures can be employed. An approach is active synchronization with randomized delays, ensuring that photon arrival times are unpredictable and decoupled from Eve’s control \cite{acheva_automated_2023}. Another method is hardware equalization, where detector efficiencies are carefully matched and stabilized to eliminate timing-dependent bias. More fundamentally, protocols such as MDI-QKD can be implemented, eliminating the need to trust the detectors entirely, making such attacks ineffective regardless of hardware imperfections \cite{ma_alternative_2012, acheva_automated_2023}.

\subsubsection{Source State Preparation Vulnerabilities}
Imperfect source state preparation in quantum communication systems can introduce vulnerabilities that compromise security. Such imperfections may arise from phase modulation errors, inconsistencies in amplitude control, or distinguishability between states in non-encoding degrees of freedom, such as temporal, spectral, or spatial modes \cite{nahar_imperfect_2023}. For instance, when optical pulses are generated by independent laser diodes, slight variations in their emission characteristics such as, wavelength distribution, pulse shape, or timing, can make them partially distinguishable. This distinguishability provides an adversary with side-channel information that could be exploited to gain knowledge about the encoded quantum states without introducing detectable disturbances. Apart from this, optical components may have wavelength-dependent behaviors that create exploitable correlations, or during system calibration, Eve may bias settings (e.g., detector efficiencies) to favor her attack. High-power laser damage attacks can also disable security-critical components, such as attenuators or monitors. 

To mitigate these risks, several countermeasures can be employed. Huang \emph{et al.} studied experimental ways of generating decoy states and proposed a calibration method to restore security against weak coherent source \cite{huang_quantum_2018}. Jiang \emph{et al.} suggested a method to treat imperfect quantum states as if they were perfect ones, by checking the likelihood of each state being in a vacuum, so that QKD stays secure even with side-channel flaws.  \cite{jiang_side-channel_2024}. Furthermore, various methods based on the use of a single laser source with active modulation reduce variations between pulses, while complete phase randomization ensures that any residual phase correlations cannot be exploited \cite{berra_modular_2023}. Spectral filtering can further homogenize the output by removing unwanted frequency components, and source monitoring can detect deviations from expected state preparation parameters \cite{berra_modular_2023, marcomini_characterising_2025}. Together, these measures help ensure that the prepared quantum states closely match the ideal specifications defined by the protocol, thereby preserving the intended security guarantees.  

\subsubsection{Backflash Light Attacks}
Backflash light attacks exploit a subtle vulnerability in single-photon detectors, where the process of registering a photon can inadvertently produce and emit a faint burst of light, known as “backflash.” This backflash can propagate back through the optical channel and be collected by an eavesdropper, such as Eve, who may use it to infer when a detection event has occurred. By correlating the timing of this leakage with the transmission and measurement settings, Eve can potentially extract sensitive information without directly interfering with the quantum signal, thus compromising the security of the system \cite{pinheiro_eavesdropping_2018, singh_backflash_2025}. 

To mitigate such risks, several countermeasures are employed. Optical isolators are commonly used to allow light to pass in only one direction, preventing the escape of backflash photons toward the source. Backflash suppression coatings can be applied to detector components to reduce the intensity of any emitted light at its origin. Additionally, spectral filters can be integrated into the detection path to block or attenuate the specific wavelengths associated with backflash, while still allowing the desired signal photons to be detected. Together, these methods help minimize backflash leakage, thereby preserving the confidentiality and integrity of quantum communication systems \cite{pinheiro_eavesdropping_2018}.

Modern QKD security has matured into a layered defense framework addressing vulnerabilities at the \textit{hardware}, \textit{protocol}, and \textit{classical communication} layers \cite{lydersen_hacking_2010, xu_secure_2020}.

\paragraph{Hardware}
At the hardware level, the security of QKD systems depends on the physical reliability and trustworthiness of optical components such as photon sources, detectors, and modulators. As described in Section \ref{QKD_Secure}, imperfections, including detector after-pulsing, efficiency mismatches, and source intensity fluctuations, can open side channels. Adversaries can exploit these vulnerabilities through timing-based strategies, such as early “blinding” or “faked-state” attacks \cite{lydersen_hacking_2010}. 
Modern QKD implementations employ active phase randomization and intensity monitoring in the experimental scenarios at the transmitter to mitigate eavesdropper detection. In some cases, the passive optical components are preferred to reduce the attack space of the adversary. Furthermore, a detailed discussion is given in Section \ref{QKD_Challenge}. 
\paragraph{Protocols}
Recent progress in QKD protocols has centered on reinforcing intrinsic security mechanisms to counter both quantum-specific and classical attack vectors. At the protocol layer, one of the most significant threats can be categorized into \textit{source-based} i.e., PNS, wavelength-dependent leakage \cite{brassard_limitations_2000, lutkenhaus_security_2000}, \textit{channel-based} i.e., Trojan horse, timing manipulations \cite{gisin_trojan-horse_2006,vakhitov_large_2001, xu_secure_2020}, and \textit{receiver-based} i.e., blinding, efficiency mismatch classes \cite{singh_backflash_2025}. These attacks take advantage of the gap between the ideal assumption of single-photon sources and the practical limitations of real hardware. Another critical issue is the risk of MITM attacks, where maintaining message integrity necessitates either pre-shared keys for classical information-theoretic MACs or post-quantum secure digital signatures such as lattice-based or FALCON schemes to safeguard QKD networks against both current and future quantum adversaries \cite{bibak_quantum_2021}. Complementary advances, including reference-frame-independent (RFI) protocols, finite-key composable security proofs and adaptive post-selection, have strengthened the resilience of QKD systems against phase misalignment, synchronization errors, and finite data constraints \cite{laing_reference-frame-independent_2010, wang_finite-key_2023}. Collectively, these developments have advanced QKD from a purely theoretical construct into a composable secure framework and practically deployable framework capable of withstanding realistic imperfections across both quantum and classical domains. Further refinements, such as basis choice randomization and phase randomization, help mitigate information leakage from imperfect state preparation, thereby enhancing the robustness of modern QKD implementations.

\paragraph{Classical Communication}
While the quantum channel in QKD ensures information-theoretic secrecy against eavesdropping on photon transmission, the accompanying classical communication layer remains a critical component influencing overall system security. These vulnerabilities are addressed through post-quantum cryptographic (PQC) primitives such as lattice-based and hash-based digital signatures (e.g., CRYSTALS-Dilithium, FALCON, SPHINCS+), which secure classical authentication channels against both present and future adversaries. Also, side-channel leaks, denial-of-service (DoS) vectors, and traffic pattern analysis remain open challenges, emphasizing the need for hybrid architectures that combine quantum security guarantees with robust classical cryptographic infrastructures.

This indicates that, the security of QKD has evolved from idealized theoretical constructs to rigorous frameworks that are practical, thus provide provable security. Modern QKD adopts a layered architecture that integrates physical, protocol, and classical to provide composable and operationally meaningful security. This progression marks a significant step toward realizing robust, scalable, and practically deployable quantum communication networks. However, several fundamental challenges remain, which are discussed in the following section.

\begin{table*}[htbp]
\centering
\caption{Comprehensive comparison of modern quantum key distribution protocols}
\label{tab:unified_qkd_comparison2}
\footnotesize
\begin{tabular}{|p{2.2cm}|p{1.8cm}|p{2.2cm}|p{1.9cm}|p{1.7cm}|p{2.1cm}|p{3.2cm}|}
\hline
\textbf{Protocol} & \textbf{Key Rate Scaling} & \textbf{Security Model} & \textbf{Source Type} & \textbf{Typical Distance} & \textbf{Implementation Complexity} & \textbf{Key Characteristics} \\
\hline
MDI-QKD & $R \sim \eta^2$ & Measurement-independent & WCP/Single photon & 200--400 km & High & Detector attack immunity; double transmission loss \\
\hline
DI-QKD & $R \sim \eta$ (very low) & Device-independent & Any quantum source & 10--50 km & Very high & Ultimate security; requires loophole-free Bell tests \\
\hline
TF-QKD & $R \sim \sqrt{\eta}$ & Measurement-independent & WCP with interference & $500\text{--}1000^+ \text{ km}$ & Moderate--High & Breaks PLOB limit; measurement-device independent \\
\hline
\end{tabular}
\end{table*}

\section{QKD Challenges} \label{QKD_Challenge}

Despite decades of development and numerous protocol innovations, conventional QKD schemes face several fundamental challenges that have hindered their widespread deployment and motivated the development of twin-field approaches.

\paragraph{Exponential Rate-Distance Scaling} The most fundamental limitation is the exponential decay of key rates with transmission distance. In an optical fiber with a loss coefficient $\alpha \approx 0.2$ dB/km, the transmittance follows $\eta = 10^{-\alpha L/10}$, causing key rates to drop exponentially. Even with optimal implementations reaching the PLOB bound, this scaling restricts practical QKD to metropolitan area networks, typically limiting secure links to distances under 300 km without quantum repeaters. 
The theoretical foundation of the rate-distance barrier is established through several fundamental bounds that limit the performance of any point-to-point QKD protocol. The Takeoka-Guha-Wilde bound, which preceded the more general PLOB bound, established that the secret key capacity of lossy optical channels is fundamentally limited by the channel transmittance. The PLOB bound sets the fundamental limit for the maximum amount of private states that can be transferred in QKD for a given quantum channel without the use of a repeater, mathematically expressed as \cite{pirandola_fundamental_2017, pirandola_advances_2020},
\begin{equation}
R \leq -\log_2(1 - \eta) \approx 1.44\eta
\end{equation}
Here, $\eta$ represents the channel transmittance and the secret key capacity $R$ is bounded by the transmittance $\eta$ of the optical link. In optical fiber, transmittance decays exponentially with distance as $\eta = 10^{-\alpha L/10}$, where $\alpha \approx 0.2$ dB/km is the fiber loss coefficient, as mentioned previously, and $L$ is the transmission distance. This exponential decay fundamentally limits secure communication to distances of a few hundred kilometers, even with optimal protocol implementations. The PLOB bound reflects a deeper physical constraint in conventional QKD, the key rate scales linearly with channel transmittance ($R \sim \eta$), making long-distance communication prohibitively slow without quantum repeaters. This limitation became the primary obstacle to building large-scale quantum networks and motivated the search for alternative approaches.
The combination of exponential distance scaling with linear rate dependence creates a fundamental barrier that cannot be overcome through improvements in detector efficiency or source quality alone.

For practical QKD implementations that use direct transmission between distant parties, the theoretical upper bound of the secure key rate is limited to 1.44 $\eta$ \cite{pirandola_advances_2020}. This upper bound is valid for all repeater-less QKD protocols, see Table \ref{tab:unified_qkd_comparison}. Linear scaling of key rates with channel transmittance creates a practical ceiling that severely restricts the achievable communication distances and throughput of conventional QKD systems. 

\paragraph{Environmental Sensitivity and Stability} QKD systems are inherently sensitive to environmental fluctuations that affect phase stability, polarization drift, and timing synchronization. Fiber-based implementations suffer from temperature variations, mechanical vibrations, and aging effects that degrade system performance over time. Maintaining stable operation requires active stabilization systems and frequent calibration, increasing system complexity and operational costs.

\paragraph{Limited Network Scalability} Conventional QKD protocols are fundamentally point-to-point, requiring dedicated quantum channels between each pair of communicating parties. Scaling to large networks necessitates trusted relay nodes or quantum repeaters, both of which introduce significant technical and security challenges. The absence of practical quantum repeaters has been a major bottleneck for large-scale QKD network deployment.

\paragraph{Integration with Classical Infrastructure} QKD systems require specialized hardware, including single-photon sources, detectors, and quantum-grade optical components that are expensive and incompatible with existing classical communication infrastructure. The need for separate quantum channels increases deployment costs and limits the economic viability of QKD for many applications.

These challenges collectively explain why, despite its theoretical elegance and proven security, conventional QKD has remained largely confined to specialized applications and research demonstrations. The fundamental rate-distance scaling limitation, in particular, represents a physical barrier that cannot be overcome through incremental improvements to existing protocols. This realization drove the search for fundamentally new approaches, ultimately leading to the twin-field paradigm that overcomes the repeaterless scaling limit while maintaining measurement-device independence.


\section{Twin-Field QKD}\label{TF-QKD}
The limitations discussed in previous sections motivate the transition toward next-generation QKD paradigms, such as DI-QKD, that eliminates reliance on device trustworthiness by basing security solely on violation of Bell inequalities, thereby addressing all implementation-specific side-channels in principle, and the TF-QKD, which directly tackles the rate-distance barrier, enabling secure key distribution beyond 500 km with standard optical fibers by exploiting single-photon interference between distant phase-stabilized fields and also integrates essential features from both MDI-QKD and decoy-state QKD, embodying a synthesis of robust theoretical guarantees and extended operational reach. Collectively, these advances represent a strategic direction for reconciling unconditional security with practical deployment constraints, here we address the latter in the present work.

TF-QKD introduced by Lucamarini \emph{et al.} in 2018 \cite{lucamarini_overcoming_2018}, represents a paradigm shift from conventional point-to-point quantum key distribution schemes. The revolutionary approach of the protocol involves Alice and Bob each transmitting phase-randomized WCP having distribution, $ p_n(\mu) = \mu^n e^{-\mu}/n! $ to an un-trusted intermediate node (Charlie), where single-photon interference creates quantum correlations by utilizing the detector yield model having probability, $Y_n = 1- (1-2P_{\text{dc}})(1-\eta)^n$ without revealing the photon's origin and incorporating the measurement error of dark counts $P_{\text{dc}}$. The fundamental insight underlying TF-QKD is the exploitation of single-photon interference in a double-transmission geometry. When Alice and Bob simultaneously send WCP with correlated phases to Charlie's measurement station, successful detection events correspond to scenarios where exactly one photon arrives from each sender. The interference between these indistinguishable photons projects the joint state onto a Bell-like subspace, creating correlations between Alice's and Bob's phase choices while maintaining uncertainty about which party emitted the detected photon.

\begin{figure*}
    \centering
    \includegraphics[width=.8\linewidth]{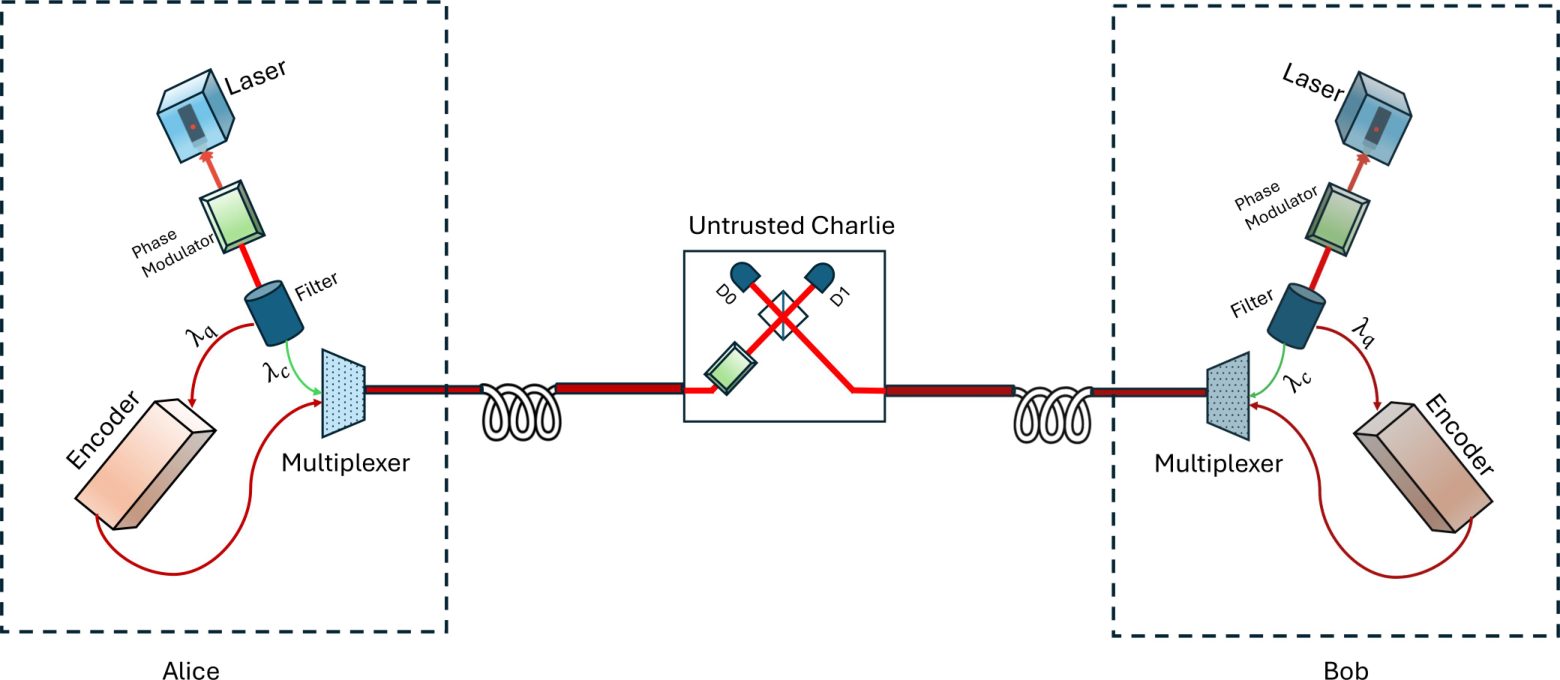}
    \caption{Schematic of a typical TF-QKD setup. Independent continuous wave lasers at Alice’s and Bob’s stations are phase-modulated, encoded ($\lambda$\textsubscript{q} for quantum and $\lambda$\textsubscript{c} for reference), spectrally filtered, and combined via wavelength-division multiplexers. The multiplexed signals traverse long‐distance fiber links to an untrusted middle node (Charlie), where they interfere on a beam-splitter and are detected by single‐photon detectors D\textsubscript{0} and D\textsubscript{1}. Successful interference events are used to generate a secure key, while the dashed boxes delineate the local preparation modules at Alice and Bob.}
    \label{fig:tfqkd_schematic}
\end{figure*}
Fig. \ref{fig:tfqkd_schematic} illustrates a typical TF-QKD setup. It can be observed that two distant parties, Alice and Bob, each prepare WCP using independent laser sources, modulated in phase and filtered. A multiplexer combines the signal ($\lambda_q$) and reference ($\lambda_c$) channels. These pulses are transmitted through optical fiber to an untrusted middle node (Charlie), where interference occurs at a beam splitter. The outputs are monitored by single-photon detectors $D_0$ and $D_1$. Successful detection events are publicly announced by Charlie, but do not reveal the key, ensuring security even if Charlie is untrusted. This setup enables long-distance QKD beyond the rate-loss limit of point-to-point QKD systems.

This geometry effectively \emph{unfolds} the traditional point-to-point channel of length $L$ into two parallel channels of length $L/2$, fundamentally altering the scaling relationship between key rate and distance. Instead of conventional linear scaling $R \sim \eta(L)$ that characterizes all repeater-less QKD protocols, TF-QKD achieves breakthrough scaling $R \sim \sqrt{\eta(L)}$, where $\eta(L) = 10^{-\alpha L/10}$ represents end-to-end channel transmittance. This square-root scaling effectively circumvents the repeater-less PLOB bound \cite{pirandola_fundamental_2017}, extending the reach of secure quantum communication by orders of magnitude.
The protocol maintains measurement-device independence, inheriting the security advantages of MDI-QKD while overcoming its quadratic rate scaling limitation. Charlie's measurement apparatus can be fully untrusted, eliminating all detector-side vulnerabilities that plague conventional QKD implementations. This combination of enhanced security and superior rate-distance scaling positions TF-QKD as a leading candidate for long-distance quantum network backbones.

To contextualize TF-QKD's breakthrough performance, Table \ref{tab:unified_qkd_comparison} provides a comprehensive comparison with established quantum key distribution approaches, highlighting the fundamental trade-offs between security assumptions, achievable distances, and implementation complexity.

The comparison reveals the unique position of TF-QKD in the QKD landscape. While conventional protocols like BB84 with decoy states offer simplicity and moderate complexity, they remain fundamentally limited by linear rate scaling. Entanglement-based protocols (E91) provide conceptual elegance, but face practical challenges in generating and distributing high-quality entangled states over long distances.

Continuous-variable QKD excels in compatibility with classical telecommunication infrastructure but typically achieves shorter distances due to finite squeezing and detection efficiency limitations. MDI-QKD successfully eliminates detector vulnerabilities but suffers from quadratic scaling due to double transmission losses, making it less competitive than TF-QKD at long distances despite its security advantages \cite{bertaina_phase_2024}.

Device-Independent QKD represents the ultimate security paradigm but requires extremely demanding experimental conditions for loophole-free Bell tests, resulting in very low key rates and a limited practical range. The stringent requirements for high detection efficiency, fast switching, and precise timing make DI-QKD primarily of theoretical interest in the near term.

TF-QKD strikes an optimal balance by combining the detector security of MDI-QKD with fundamentally superior rate-distance scaling. While the protocol requires moderate to high implementation complexity due to phase stabilization and interference requirements, these technical challenges are well within the capabilities of current photonic technology. The resulting combination of security, performance, and practicality makes TF-QKD the clear choice for intercity and potentially intercontinental quantum communication links.

Importantly, TF-QKD's advantages become increasingly pronounced at longer distances, where the square-root scaling i.e. $R \sim \sqrt{\eta}$, dramatically outperforms all conventional approaches by leveraging single-photon interference between distant phase-locked weak coherent pulses. This shows that a practical long-distance QKD is achievable using existing infrastructure while maintaining the independence of the measurement-devices and counter various practical vulnerabilities. These characteristic position the TF-QKD protocol as an enabling technology for large-scale quantum networks and the future quantum internet backbone, where secure communication over thousands of kilometers is essential for practical deployment.

\section{TF-QKD Characterization}
The mathematical characterization of QKD protocols is fundamentally based on quantifying the relationship between resource consumption and cryptographic output. The traditional approach has focused primarily on the secret key rate $R$, defined by the asymptotical formula expressed through Shannon entropy functions that account for QBER and the costs of error correction and privacy amplification procedures. However, the conventional TF-QKD framework presents a comprehensive mathematical treatment through the concept of total efficiency, which incorporates both quantum and classical resource utilization, leading to a revised total key rate $R$ that accounts for the joint pulse repetition frequencies of quantum and classical communication channels \cite{bebrov_relation_2024}.
 
\subsection{Probability of Click}
In TF-QKD, the central untrusted measurement station, referred to as Charlie, plays a crucial role in establishing correlations between Alice and Bob’s quantum signals. At this node, the WCP sent by the two users interfere at a 50:50 beam splitter, and the detection outcome is determined by the clicks registered on Charlie’s single-photon detectors. The click probability at Charlie's detector quantifies the likelihood that a detection event occurs during each interference trial, which is given as, 
\begin{equation}
    Q_\mu = \sum_{n=0}^\infty p_n(\mu) Y_n
\end{equation}
In TF-QKD, since both Alice and Bob send WCP toward a central node (Charlie), the corresponding gain expression involves two photon-number distributions as, 
\begin{equation}
Q_{\mu, \nu} = \sum_{n=0}^{\infty} \sum_{m=0}^{\infty} p_n(\mu) \, p_m(\nu) \, Y_{n,m}
\end{equation}
where \( p_n(\mu) \) and  \( p_m(\nu) \) are the Poisson probabilities of sending \( n \) and \( m \) photons by Alice and Bob, respectively, and \( Y_{n,m} \) is the joint yield.
\subsection{Error characterization}
Depending on practical factors such as optical channel loss, pulse intensity, detector efficiency, dark counts, and phase stability. Accurate modeling of this probability is essential for estimating the secure key rate and assessing the system’s overall performance. 
\begin{itemize}
    \item Photon-number mismatch error contribution:
    Alice and Bob send random signal to Charlie's node, and the probability of a mismatch is characterized as,
    \begin{equation}
        E_\mu = \frac{1}{Q_\mu}\sum_{\substack{n,m \\ n \neq m}} p_{n,m}Y_{n,m}e_{n,m}
    \end{equation}
    Here, $p_{n,m} = p_n p_m$, where $n$ and $m$ are the mean photon number of Alice and Bob respectively.
    
    \item Phase Slicing Error:
     Alice and Bob prepare coherent states with random global phases. To enable interference, they divide the phase space $ [0,2\pi] $ into $M$ equal slices having width, $\Delta \phi = 2\pi/M$ and after transmission, they each announce which slice their phase fell into. Here, a mismatch arises due to randomness known as the phase slicing error. It is a real and important error source in TF-QKD. It affects the interference quality between Alice and Bob’s pulses at Charlie, and thus contributes to the QBER as,
    \begin{equation}
        E_M = \frac{1}{2} (1 - \frac{M}{2\pi}\sin{(2\pi/M)})
    \end{equation}
    Here, even though they’re in the same slice, the actual phase difference.
    \item Decoy State Errors:
    In TF-QKD, Alice and Bob each send phase-randomized WCPs to a central untrusted node (Charlie). These pulses are not perfect single photons; they follow a Poisson distribution, so there is a non-zero chance that some pulses contain more than one photon. These multi-photon components are vulnerable to PNS attacks by an eavesdropper. To estimate the contribution from single-photon events, TF-QKD uses the decoy-state method: each user randomly varies the intensity of their pulses (e.g., signal, decoy, and vacuum states). These variations allow for statistical estimation of the single-photon yield and error rate, which are necessary to bound Eve's information.

    \begin{equation}
    R(\mu, L) = Q_1^{\mu, L} \left[ 1 - h\left(e_1^{\mu, L}\right) \right] - f \, Q_\mu^L \, h\left(E_\mu^L\right)
    \end{equation}
    Here, \( Q_1^{\mu, L} \) is Lower bound on single-photon gain at intensity \( \mu \), \( e_1^{\mu, L} \) is the upper bound on the error rate of single-photon events, while \( Q_\mu^L \) represents the overall gain (detection probability) at signal intensity \( \mu \) and \( E_\mu^L \) is the  QBER at intensity multiplied to error correction efficiency \( f \). The function, \( H(x) \) is the binary entropy function, defined as \( H(x) = -x \log_2 x - (1 - x) \log_2(1 - x) \).
\end{itemize}
\subsection{Asymptotic Limit}
The asymptotic regime assumes that the communicating parties share infinitely many quantum signals during the QKD process. Under this assumption, statistical fluctuations become negligible, and the secret key rate can be expressed in a compact analytical form. Such asymptotic formulas provide a theoretical benchmark in scaling behavior and guiding the evaluation of practical implementations. The corresponding asymptotic formula for TF-QKD is given as follows,
\begin{equation}
    R = \frac{d}{M} \Bigg[ Q_1(1 - H(\xi_1)) - f\times Q_\mu H(E) \Bigg]
\end{equation}
Here, $Q_1$ is the gain of single photon events, which are actual signals, $\xi_1$ is the single photon error, $E = E_\mu + E_M -2 E_\mu E_M$ is the total error due to misalignment of phase and slicing, and the overall asymptotic Formula is given by the decoy state standard expression, where $H$ is the binary entropy function.

\begin{figure}
    \centering
    \includegraphics[width=1\linewidth]{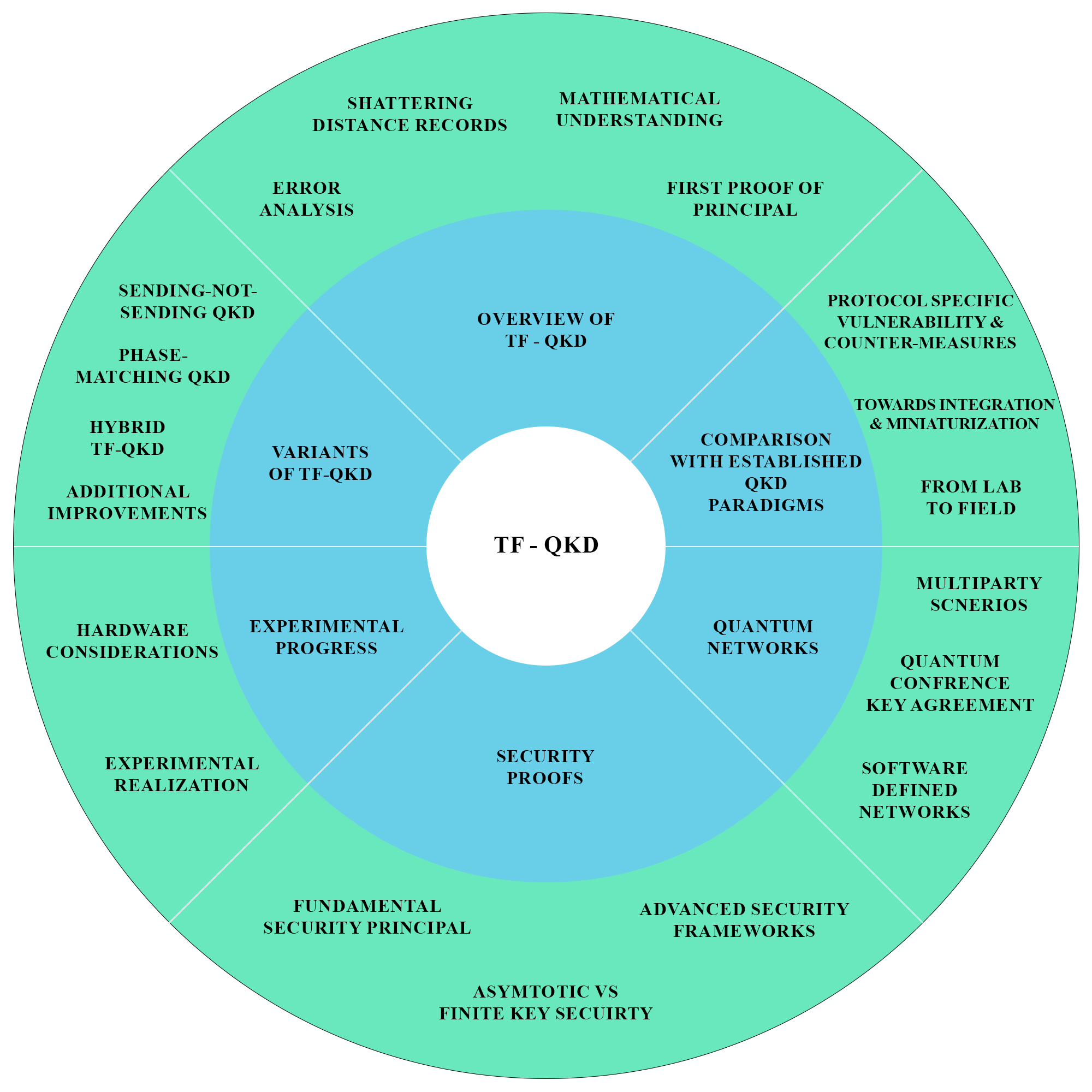}
    \caption{Illustration of the conceptual landscape of TF-QKD, where the central core highlights the three main areas of research and the surrounding areas represent key subtopics.}
    \label{fig:TFQKD_framework}
\end{figure}
The mathematical characterization of TF-QKD provides a comprehensive framework that encompasses the interference-based detection, decoy-state estimation, and error modeling such as phase slicing, photon-number mismatch, which together determine the security and efficiency of TF-QKD. Further, different variants of TF-QKD protocol have been constructed for better scaling and feasibility under realistic conditions.

\section{TF-QKD Variants} \label{TF-Var}
Broadly, we classify the TF-QKD into three major families: Phase-Matching QKD (PM-QKD), Sending-or-Not-Sending Protocol (SNS-QKD) and Hybrid TF-QKD, which are divided into various schemes and variants discussed as subcategories, as shown in Table \ref{tab:comprehensive_tfqkd}. 
\begin{table*}[htbp]
\centering
\caption{Comprehensive timeline of TF-QKD developments: protocols, security frameworks, and experimental milestones (2018--2025)}
\label{tab:tfqkd_comprehensive_timeline}
\adjustbox{width=\textwidth,center}{
\footnotesize
\begin{tabular}{|p{1.2cm}|p{3.2cm}|p{2.8cm}|p{2.5cm}|p{6.5cm}|}
\hline
\textbf{Year} & \textbf{Reference} & \textbf{Protocol/Framework} & \textbf{Category} & \textbf{Key Contributions and Achievements} \\
\hline
\multicolumn{5}{|c|}{\textbf{Foundation Era (2018)}} \\
\hline
\multirow{2}{*}{2018} & Lucamarini \emph{et al.} \cite{lucamarini_overcoming_2018} & Original TF-QKD & Protocol \& Theory & Introduced twin-field concept; achieved $\sqrt{\eta}$ scaling; broke repeaterless PLOB bound; demonstrated feasibility with phase-randomized coherent states \\
\cline{2-5}
& Ma \emph{et al.} \cite{ma_phase-matching_2018} & PM-TF-QKD & Protocol \& Security & Generalized TF-QKD via phase-matching; composable security proof using optical-mode entanglement; introduced phase post-compensation \\
\hline
\multicolumn{5}{|c|}{\textbf{Protocol Development Era (2019--2022)}} \\
\hline
\multirow{3}{*}{2019} & Wang \emph{et al.} \cite{wang_beating_2019} & SNS-QKD & Protocol \& Experiment & Sending-or-not-sending protocol; avoided phase post-selection; 421 km demonstration; improved phase misalignment tolerance \\
\cline{2-5} 
& Zhou \emph{et al.} \cite{zhou_asymmetric_2019}
& No-Post Selection TF-QKD & Protocol \& Security & Provided a new security proof countering collective and coherent attacks in the asymptotic case \\
\cline{2-5}
& Cui \emph{et al.} \cite{cui_twin-field_2019} & No-Post Selection TF-QKD & Protocol \& Security & Eliminated phase post-selection while retaining $\sqrt{\eta}$ scaling; simpler implementation with higher key rates \\
\cline{2-5}
& Grasselli \& Curty \cite{grasselli_alvaro_navarrete_and_marcos_curty_asymmetric_2019} & Decoy-state TF-QKD & Security Framework & Practical decoy-state analysis; showed two decoys sufficient to beat repeaterless bounds; tight parameter estimation \\
\cline{2-5}
& Yin \emph{et al.} \cite{yin_finite-key_2019} & Finite-key TF-QKD & Security Framework & First finite-key security proof using entropic uncertainty relations; demonstrated practical security over 800 km \\
\hline
\multirow{2}{*}{2021} & Curras-Lorenzo \emph{et al.} \cite{lorenzo_finite-key_2021} & Finite-key analysis & Security Framework & Tight finite-key analysis for no-post-selection TF-QKD; proved $\sqrt{\eta}$ scaling persists in finite-size regime \\
\cline{2-5}
& Pittaluga \emph{et al.} \cite{pittaluga_600-km_2021} & Dual-band stabilization & Experiment & 600 km repeater-like communication with advanced phase stabilization techniques \\
\hline
\multirow{2}{*}{2022} & Zhou \emph{et al.} \cite{zhou_twin-field_2022} & Partial post-selection & Protocol \& Security & Hybrid protocol combining post-selection modes; finite-key secure with improved rates; universal composability \\
\hline
\multicolumn{5}{|c|}{\textbf{Record-Breaking Era (2023--2025)}} \\
\hline
\multirow{2}{*}{2023} & Liu \emph{et al.} \cite{liu_1002_2023} & 1002 km demonstration & Experiment & World record distance with finite-key analysis; solidified theoretical predictions with practical validation \\
\hline
2024 & Han Du \emph{et al.} \cite{du_twin-field_2024} & Photonic chip integration & Technology & Optical injection locking on-chip; path toward miniaturized TF-QKD systems \\
\cline{2-5}
& Chen \emph{et al.} \cite{chen_twin-field_2024} & Practical stabilization & Experiment & 502 km with local frequency reference; practical phase stabilization methods for field deployment \\
\hline
\multirow{3}{*}{2025} 
& Wang \emph{et al.} \cite{zhan_experimental_2025} & Long-distance demo & Experiment & 830 km fiber transmission; advanced detector technologies; field deployment validation \\
\hline
\end{tabular}
}
\end{table*}

\subsection{Phase-Matching TF-QKD}

PM-QKD first formalized in the original Phase-Matching Quantum Key Distribution proposal, is a coherent-state, single photon-interference realization of the twin-field concept that achieves the characteristic $R\sim O(\sqrt{\eta})$ scaling without trusted relays, while retaining measurement-device-independence. In this model, Alice and Bob encode key bits in the global phase of weak coherent states $ \ket{\sqrt{\mu}e^{i\theta_a}} $ and $ \ket{\sqrt{\mu}e^{i\theta_b}} $, transmit them to an untrusted interferometer, and determine the bit value from which the output detector clicks, keeping only rounds where the phases lie in the same post-selected slice \cite{ma_phase-matching_2018}. This combination of phase-slice post-selection and decoy-state estimation removes the need for active per-pulse phase locking while preserving the $\sqrt{\eta}$ advantage that enables performance beyond the repeaterless PLOB bound. The approach has been consistently interpreted within the measurement-device-independent framework \cite{lin_simple_2018}, where the protocol is seen as an MDI-QKD subclass relying on single-photon detection rather than coincidence counts, explaining the rate-loss scaling improvement through the same entanglement-based reduction that underpins standard MDI security.

Two engineering mechanisms are central to making PM-QKD field-deployable. First, phase post-compensation, implemented by publicly disclosing coarse-grained phase-slice indices and retaining only matched-slice events, tolerates slow phase drift and relaxes stringent locking requirements \cite{ma_phase-matching_2018, yu_free-space_2021}. Second, multi-intensity decoy states bound the single-photon contribution and phase error rate in the presence of channel loss and noise, ensuring that the security proof, whether in continuous or discrete phase settings, holds in practice \cite{shao_phase-matching_2023, zhang_phase-matching_2021}. These techniques allow standard telecom hardware to realize the theoretical performance advantage. Robustness to realistic source imperfections has been addressed through explicit finite-size, decoy-state analyses that model intensity fluctuations and imperfect phase modulation \cite{yu_decoy-state_2021}. This conservatively bounds single-photon yields and phase errors even when preparation deviates from the ideal, paralleling the source-flaw-tolerant methods in related discrete-phase implementations. Discrete phase randomization itself is a recurring pragmatic modification, with security proven against unambiguous state discrimination and photon-number-splitting strategies, and simulation results showing that modest phase sets recover near-continuous performance \cite{zhang_phase-matching_2021}. These findings align with the simplifications suggested in the MDI-based security re-derivation, where the symmetry arguments link phase error directly to observed bit error without the heavy combinatorics of earlier proofs.

Beyond fiber deployments, PM-QKD has been adapted to free-space and asymmetric channels, such as those in satellite-ground or airborne links, where turbulence leads to random, unequal transmittance in the two arms. A generalized asymmetric-channel formulation using four-intensity decoys and phase-slice post-selection has been shown to maintain positive key rates under log-normal turbulence models \cite{yu_free-space_2021, cui_satellite-based_2022}, consistent with the asymmetric-intensity variants anticipated in the simplified security framework. These studies confirm the feasibility of long slant paths, provided phase-matching stability can be maintained over atmospheric fluctuations.

Alternative source types further diversify the PM-QKD landscape. Replacing coherent states with single-photon entanglement removes multi-photon components and, in principle, eliminates the need for decoys \cite{li_phase_2019, li_extended_2022}. Here, Alice and Bob each prepare an entangled state between a local mode and a traveling mode, sending the latter to interfere at the central station. Security derives from the absence of multi-photon pulses, and simulations show superior finite-key performance compared to WCP implementations, especially when channel loss is high or data blocks are short. An extended SPE scheme incorporates delocalized-state measurements and can be interpreted as a TF-QKD link joined by an SPE-based repeater, delivering TF/PM-like key rates with improved reach. Finally, integrating quantum memories into the phase-matching architecture has been proposed to further reshape rate-loss scaling (Memory-Assisted PM-QKD). By storing one arm and interfering after successful retrieval, the per-channel-use scaling can, under idealized conditions, improve from $ O(\sqrt{\eta_{tot}}) $ to $ O(2^{n-1}\sqrt{\eta_{tot}}) $ with $n-1$ memory stations. Even with realistic dephasing and dark counts, simulations predict achievable distances beyond 700 km with high-quality memories. This "repeater-lean" use of memories positions PM-QKD not just as a repeater-less solution, but as a natural backbone for incremental repeater-assisted upgrades.

Overall, PM-QKD provides a versatile platform adaptable to fiber, free-space, and satellite links. It combines the fundamental rate--distance scaling advantage of Twin-Field protocols with practical engineering solutions such as advanced photon sources, memory assistance, and robust error handling, making it a strong candidate for next-generation large-scale quantum communication networks. By operating within predefined phase slices, PM-QKD avoids complications associated with post-selection and the need for separate estimation procedures, common in earlier TF-QKD variants. This design simplifies analysis and yields tighter key rate bounds, particularly in the finite-key regime, where statistical fluctuations are significant. The phase-sifting method also enhances parameter estimation efficiency, contributing to improved key rates at long distances. Furthermore, PM-QKD offers a clean framework for composable security proofs by modularizing the encoding, sifting, and key distillation stages. This modularity facilitates security analysis against general attacks and makes the protocol more amenable to integration into larger cryptographic systems that require composable security. As a result, PM-QKD has become a reference model in the study of Twin-Field QKD, with many subsequent variants adopting its phase-sifting technique or building on its security framework. One such evolution is the Sending-or-Not-Sending Twin-Field QKD (SNS-QKD) protocol, which introduces a different strategy for key generation and error handling to further optimize performance in practical deployments.

\subsection{Sending-or-Not-Sending Protocol TF-QKD}

The original SNS-QKD, proposed by Yu \emph{et al.} (2019), refines the TF-QKD framework by introducing a novel probabilistic sending mechanism \cite{yu_sending-or-not-sending_2019}. In each time slot, Alice and Bob randomly decide whether to send a phase-randomized WCP (“send”) or emit nothing (“not-send”). Key bits are generated from the interference outcomes at the untrusted measurement node when one party sends and the other remains silent, which produces a predictable detection pattern. Cases where both parties either send or both do not send are used to estimate channel parameters and detect eavesdropping. This selective use of events allows SNS-QKD to sidestep the need for phase matching or active phase slicing, which greatly reduces the experimental complexity and sensitivity to phase drift. The main strength of SNS-QKD is its robustness against optical noise, misalignment, and phase instability, making it particularly suited for long-distance fiber-based implementations. Since the protocol relies only on the relative probabilities of sending versus not-sending, it can tolerate more substantial fluctuations in the optical channel compared to phase-sensitive protocols. Moreover, SNS-QKD integrates naturally with the decoy-state method, using different intensities for “send” pulses to securely estimate single-photon contributions, a critical aspect for defending against PNS attacks \cite{yu_sending-or-not-sending_2019}. 

Early theoretical work introduced the decoy-state method for SNS-TF QKD under measurement imperfections. The SNS-QKD encodes key bits by probabilistically sending or not sending light in the $Z$ basis while reserving single-photon interference for the $X$ basis; this small design choice removes phase-announcement issues in key generation, lets the standard decoy-state method apply cleanly, and makes the protocol strikingly tolerant to misalignment compared with early TF-QKD ideas \cite{shan_sending-or-not-sending_2024, wang_twin-field_2018}. The first experiment validated the concept at 300 km with a secure rate of $2.65\times 10^{-6}$ bits/pulse, clearly surpassing PLOB and establishing the standard workflow of phase post-selection in $X$, single-click sifting, and decoy-state estimation \cite{liu_experimental_2019}. Subsequent finite-key demonstrations scaled this to 502 km with full fluctuation accounting, Chernoff analysis over $2.88\times 10^{11}$ pulses, $M=16$ phase slices, closing the gap between asymptotic theory and practice and showing that SNS-QKD's $\sqrt{\eta}$ scaling survives realistic data volumes \cite{liu_1002_2023}. Pushing distance further, a 658 km field system combined SNS-QKD with distributed vibration sensing on the same fiber, an early example of multifunctional quantum infrastructure, while still beating PLOB in the finite-key regime \cite{chen_quantum_2022}. Another milestone removed any shared-laser requirement: a 509km secure link with truly independent lasers used reference-pulse phase estimation and postprocessing, proving SNS-QKD can forgo tight phase locking without sacrificing security \cite{chen_sending-or-not-sending_2020}. A complementary architecture brought "plug-and-play" convenience, Charlie's common source and reciprocal paths passively cancel phase drift, provided the usual inbound-light monitoring to counter Trojan-horse risks \cite{xue_plug-and-play_2021}. To simplify phase/frequency management in long hauls, local-reference designs showed that high-visibility interference can be achieved without distributing an optical reference over the span \cite{chen_twin-field_2024}.

Security theory has kept pace with these engineering advances. Composable, unconditional proofs for finite pulses established tight single-photon and phase-error bounds across realistic block sizes $10^{10}-10^{12}$, providing the backbone for modern deployments \cite{jiang_unconditional_2019}. Efficient treatments of side-channel leakage generalized the tagged-state method and finite-key estimation so that small basis-dependent leaks (from modulators, spectrum, timing) can be quantified and tolerated with optimized parameters \cite{ma_efficient_2025, lu_practical_2021, lu_sending_2021}. Source imperfections once regarded as secondary are now incorporated explicitly into the security framework: imperfect vacuum (finite extinction) in the "not-sending" operation, discrete rather than continuous phase randomization, and weak randomness in the choice variables all admit secure bounds with modest rate/distance penalties when properly modeled \cite{hu_sending-or-not-sending_2022, jiang_sending-or-not-sending_2020, jiang_security_2022}. Finite-key analyses that explicitly include intensity fluctuations show how a pre-characterized stability window $\leq 2\%$ at 300 km keeps the key-rate loss to a few percent while retaining advantage over PLOB \cite{lu_finite-key_2021}. A universal framework binds these strands together, offering a single composable analysis umbrella that covers baseline SNS, robust/post-selected variants, and engineering tweaks while supporting joint parameter optimization across intensities, slice widths, and probabilities \cite{hu_universal_2022}.

On the imperfections front, SNS-QKD's signature robustness is now quantified and engineered. Because $Z$-basis bits do not require interference, large misalignment can be pushed into the $X$ basis and safely estimated there; theory shows a positive key well beyond 500--700 km even with double-digit misalignment, and tolerable thresholds far above traditional decoy MDI-QKD \cite{wang_twin-field_2018}. Practical measurement asymmetries, imbalanced beam splitters, detector-efficiency mismatch, and dark-count asymmetry, were folded into analytic gain/QBER models, revealing, for example, that realistic beam-splitter deviations and detector dark-count skews drive nontrivial rate and distance shifts at $>$300 km and should be explicitly calibrated rather than assumed symmetric \cite{sun_sending-or-not-sending_2022}. To harden real systems against drifting optics and environments, redundant-space encoding (time/polarization/wavelength diversity) raises the odds that at least one mode sits in a low-noise slice, delivering sizable rate gains in high-loss or high-phase-noise regimes with modest system complexity \cite{xu_sending-or-not-sending_2023}. For highly dynamic links, a zigzag decoy schedule alternates intensity sets over time blocks to sharpen single-photon/yield estimates in finite keys, boosting rates and adding distance headroom without new hardware \cite{jiang_zigzag_2020}.

Two complementary families of countermeasures target practical leakage and source trust. On the spectral side, frequency-tag correlations between settings and emitted spectra were quantified; the resulting bounds (via distinguishability parameters) translate directly into design targets for filtering and linewidth control to keep rate loss below a few percent at 400 km \cite{lu_practical_2021}. In parallel, light-source monitoring both passive taps and more features complete modules, that let the implementations measure photon-number statistics, spectrum, timing, and vacuum leakage in real time; the data folds back into decoy and sifting to filter anomalous pulses with only percent-level rate overhead, while raising the bar against Trojan-horse and modulator-leak attacks \cite{qian_passive_2022, qiao_sending-or-not-sending_2020}. Analyses that jointly treat flawed (e.g., amplitude/phase errors) and leaky sources show secure operating regions and trade-offs among extinction, isolation, and leakage coefficients, giving concrete engineering tolerances for long-distance operation \cite{lu_sending_2021}.

Deployment-oriented work addresses the messy realities of networks. Asymmetric links ubiquitous in real topologies cannot simply be fixed by padding loss without sacrificing the near-end's advantage. Theory and practice now support asymmetric SNS-QKD via per-party intensity/probability optimization and modified decoy equations, recovering large fractions of symmetric performance and, in some cases, extending distance compared with naive attenuation balancing \cite{zhou_asymmetric_2019}. Metropolitan field trials demonstrated multi-day stability with automated polarization/phase calibration, intensity bias tracking, and slice-wise post-selection, all inside a full finite-key treatment and with QBER drift contained to a couple of percent \cite{yu_sending-or-not-sending_2019}. Beyond fiber, free-space adaptations employ channel-aware real-time acceptance and predefined-threshold selection to discard low-transmittance slots before detection, improving SNR and trimming QBER under turbulence, an important step toward satellite and hybrid links while preserving the SNS-QKD distance benefit \cite{yu_prefixed-threshold_2022}.

Together, these results form a coherent picture of SNS-QKD began as a security-clean reformulation of TF-QKD that relocates phase sensitivity to where it can be estimated i.e. $X$-basis and keeps key generation interference-free in $Z$-basis \cite{shan_sending-or-not-sending_2024, wang_twin-field_2018}; it was then proven secure with finite pulses and field-validated from 300 km to beyond $600$ km with increasingly practical stabilization and even independent lasers \cite{liu_experimental_2019, liu_1002_2023, chen_quantum_2022, chen_sending-or-not-sending_2020}. Along the way, careful modeling of measurement asymmetries and source non-idealities prevented optimism bias in simulations and experiments \cite{sun_sending-or-not-sending_2022, hu_sending-or-not-sending_2022, jiang_sending-or-not-sending_2020}, while monitoring and leakage-aware proofs closed realistic side channels \cite{ma_efficient_2025, qian_passive_2022, lu_practical_2021, lu_sending_2021, qiao_sending-or-not-sending_2020}. Finally, engineering patterns local frequency references, plug-and-play reciprocity, redundancy, asymmetric optimization, zigzag scheduling, and free-space real-time selection make the protocol resilient across deployment scenarios without undermining composable security \cite{chen_twin-field_2024, xue_plug-and-play_2021, xu_sending-or-not-sending_2023, zhou_asymmetric_2019, jiang_zigzag_2020, yu_prefixed-threshold_2022}. In short, SNS-QKD has matured into a protocol family with deep security foundations and a growing catalog of practical techniques that collectively enable ultra-long-distance, repeater-less QKD under real-world constraints. Owing to its favorable balance between theoretical rigor, experimental feasibility, and strong noise tolerance, SNS-TF QKD has become one of the most prominent and widely adopted TF-QKD variants. These characteristics naturally motivate further exploration into hybridized schemes that combine the strengths of SNS-TF QKD with other TF-QKD variants, as discussed in the following subsection on Hybrid TF-QKD.




\begin{table*}[htbp]
\centering
\caption{Comparison of TF-QKD variants}
\label{tab:unified_qkd_comparison}
\footnotesize
\begin{tabular}{|p{2.2cm}|p{1.8cm}|p{2.2cm}|p{1.9cm}|p{1.7cm}|p{2.1cm}|p{3.2cm}|}
\hline
\textbf{Protocol} & \textbf{Key Rate Scaling} & \textbf{Security Model} & \textbf{Source Type} & \textbf{Typical Distance} & \textbf{Implementation Complexity} & \textbf{Key Characteristics} \\
\hline
\hline
PM-QKD & $R \sim \sqrt{\eta}$ & Measurement-independent & WCP with phase matching & 400--800 km & Moderate--High & Phase post-compensation; simplified analysis \\
\hline
SNS-QKD & $R \sim \sqrt{\eta}$ & Measurement-independent & WCP (sending-or-not) & 500--1000 km & High & No phase post-selection; enhanced security \\
\hline
No-postselection TF-QKD & $R \sim \sqrt{\eta}$ & Measurement-independent & WCP with active compensation & 400--900 km & High & Eliminates post-selection; improved efficiency \\ \hline
\end{tabular}
\label{tab:TFonly}
\end{table*}

\subsection{Hybrid TF-QKD}

Hybrid variants of TF-QKD integrate the core advantages of single-photon interference and $\sqrt{\eta}$ scaling with features from other QKD paradigms, implementation optimizations, or advanced security techniques. One prominent line of development replaces continuous phase randomization with a finite discrete set of phases, easing hardware demands and improving stability. In these discrete-phase TF-QKD protocols \cite{li_alternative_2023, zhang_twin-field_2020}, Alice and Bob select from $M$ equally spaced phases before interference, and for $M \geq 16$, the performance approaches that of continuous-phase TF-QKD while simplifying modulators and increasing tolerance to phase drift. Alternative discrete-phase schemes \cite{li_alternative_2023} extend this approach to central-source and double-pass architectures, blending discrete-phase encoding with plug-and-play or self-compensating designs, and fully discrete-phase implementations \cite{lorenzo_twin-field_2021} demonstrate that practical performance can match continuous-phase designs under realistic noise, making these hybrids especially suitable for field deployment.

A closely related class incorporates double-pass (plug-and-play) transmission, where pulses travel forward and backward along the same fiber to naturally compensate for phase fluctuations. Double-pass measurement-device-independent TF-QKD (DP-MDI-TF-QKD) \cite{yin_measurement-device-independent_2019} merges TF-QKD’s distance scaling, MDI-QKD's immunity to detecting side-channel attacks, and double-pass phase self-compensation, avoiding the need for active phase tracking. Discrete-phase double-pass TF-QKD variants \cite{li_alternative_2023} combine these benefits with reduced hardware complexity, offering high performance over long distances and making them promising for robust field-deployable systems where phase drift and calibration overhead are critical limitations. Some designs focus on source-side improvements, replacing WCP with specialized photon sources to improve photon statistics. In particular, the use of heralded single-photon sources (HSPS) \cite{liu_efficient_2022} reduces multi-photon probability, mitigating PNS attacks while retaining TF-QKD’s scaling. Heralding increases sifting efficiency by pre-selecting valid time slots, which is especially beneficial in high-loss regimes.

Phase post-selection strategies have also been modified to improve efficiency and phase noise tolerance. Partial phase post-selection \cite{zhou_twin-field_2022} accepts detection events within a predefined phase mismatch window, reducing sifting loss compared to full post-selection and relaxing phase stabilization requirements. In contrast, no-phase-post-selection TF-QKD \cite{lei_twin-field_2025} removes filtering entirely, retaining all detection events and employing enhanced parameter estimation to maintain security. This raises raw key rates and reduces synchronization demands, although it increases error correction complexity.

Robustness to reference frame instability and source flaws has been addressed by adapting concepts from reference-frame-independent QKD. Reference-frame-independent TF-QKD \cite{sekga_reference_2021} with source flaws estimates channel parameters without fixed phase alignment, tolerating significant hardware imperfections and enabling operation without long-term phase locking. Another direction is integration with classical optical networks. Coherent-based TF-QKD \cite{chan_security_2022} over classical optical fiber links couples quantum transmission with optical-layer encryption, supporting quantum-classical co-propagation. Next-generation coherent-detection TF-QKD \cite{chan_security_2021} and national-scale telecommunication integration \cite{pittaluga_coherent_2024} exploit hybrid detection and multiplexing schemes to allow TF-QKD to share infrastructure with DWDM traffic. Some protocols address realistic finite-size effects by merging TF-QKD with advanced statistical methods. Repeater-less QKD with efficient finite-key analysis \cite{maeda_repeaterless_2019} combines TF-QKD’s interference model with tight entropy bounds, enabling secure key generation above the PLOB bound with practical block sizes of $10^{7}$-$10^{8}$ signals. Three-pulse differential phase encoding TF-QKD \cite{sharma_twin_2024} also fits this category, blending TF-QKD with multi-pulse differential phase encoding to improve robustness to phase noise and enhance spectral efficiency.

In general, these hybrid TF-QKD designs address critical challenges in the deployment of long-distance quantum communication. Hardware simplification and stability are advanced by discrete-phase encoding \cite{lorenzo_twin-field_2021, li_alternative_2023, maeda_repeaterless_2019} and double-pass architectures \cite{yin_measurement-device-independent_2019, li_alternative_2023}; security is reinforced by MDI immunity \cite{yin_measurement-device-independent_2019}, RFI tolerance \cite{sekga_reference_2021}, and HSPS-based PNS-attack resistance \cite{liu_efficient_2022}; operational efficiency is improved through reduced post-selection losses \cite{zhou_twin-field_2022, lei_twin-field_2025} and optimized finite-key performance \cite{maeda_repeaterless_2019}; and network integration is facilitated by compatibility with classical channels \cite{chan_security_2022, chan_security_2021, pittaluga_coherent_2024}. By combining TF-QKD’s inherent distance advantage with targeted adaptations from other QKD frameworks, these variants pave the way for practical, scalable, and secure quantum networks.

\begin{figure}
    \centering
    \includegraphics[width=1\linewidth]{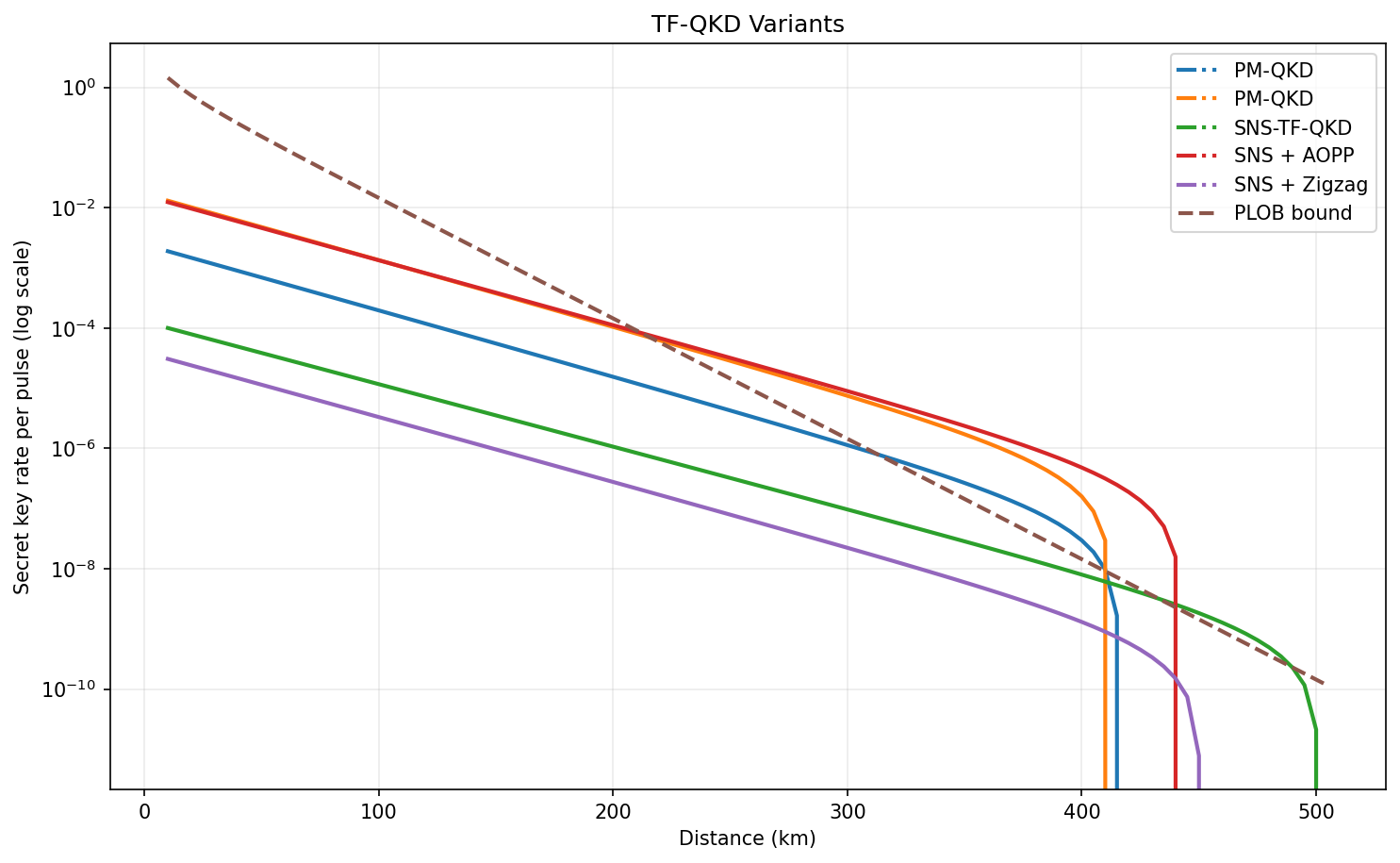}
    \caption{Comparison of secret key rate versus transmission distance for various TF-QKD schemes. The plot shows the performance of PM-TF-QKD variants \cite{ma_phase-matching_2018, lin_simple_2018} and SNS-QKD variants \cite{wang_twin-field_2018, xu_sending-or-not-sending_2023, jiang_zigzag_2020}, with the PLOB bound included as the fundamental baseline for direct transmission.}
    \label{fig:TFRKR_PLOB}
\end{figure}

\subsection{Additional Improvements}

Recent research on TF-QKD has introduced a diverse set of enhancements aimed at improving key rates, extending transmission distances, strengthening robustness, and facilitating practical deployment. A recurring theme has been protocol-physical layer integration, where cooperative frequency distribution has enhanced phase-reference stability and reduced system complexity, thereby enabling higher secure key rates over ultra-long distances \cite{meda_qkd_2022}. Alternative encoding strategies, such as three-pulse differential phase encoding, have demonstrated improved noise tolerance and lower error rates in asymmetric channels \cite{sharma_twin_2024}, while Sagnac-loop architectures have been used to mitigate back-scattering noise, achieving stable interference visibility and resilience to environmental fluctuations \cite{mandil_managing_2021}. In addition, innovations in the detection stage, such as the use of $4\times4$ beam splitter arrays, have increased parallel detection efficiency and boosted throughput without a proportional rise in hardware complexity \cite{pandey_increased_2024}.

In this context, parameter optimization remains a central focus. For example, effective parameter analysis in SNS-QKD with untrusted sources has yielded tighter bounds on key quantities, leading to more accurate yield estimation and enhanced security \cite{huang_effective_2025}. Metaheuristic methods, including particle swarm optimization, have automated parameter tuning across a wide range of distances with significant performance gains \cite{liu_twin-field_2025}. In addition, improved modeling of background photons and dark counts has informed error mitigation strategies for field-deployed systems \cite{virzi_characterizing_2023}, while machine learning techniques, such as XGBoost-assisted estimation, have reduced optimization time and enabled real-time system adaptation \cite{dong_optimization_2022}. In free-space channels, prefixed-threshold real-time selection for SNS-QKD has mitigated turbulence-induced fluctuations, ensuring stable key generation under varying atmospheric conditions \cite{yu_prefixed-threshold_2022}.

Hardware-level advances have also strengthened TF-QKD. The integration of high-precision atomic clocks has improved phase coherence across extended distances, enabling record-setting transmissions without active fiber noise compensation \cite{clivati_atomic_2023}. Entanglement-assisted TF-QKD, which employs entangled photon sources in place of independent lasers, has relaxed phase-reference stabilization requirements and achieved secure transmission beyond 600 km \cite{li_long-distance_2021}. Refined statistical fluctuation analysis has provided tighter single-photon yield and error rate bounds, particularly enhancing finite-key performance \cite{park_improved_2021}. Moreover, optimized decoy-state balances and sifting strategies have improved performance across both symmetric and asymmetric channels with reduced computational cost \cite{wang_optimized_2020}. Novel encoding approaches, including the use of orbital angular momentum (OAM) modes, have introduced multidimensional channels that increase capacity and resilience to turbulence \cite{meng_twin-field_2020}.

The SNS-QKD family has seen further refinements, such as zigzag parameter variation to mitigate finite-key effects without additional pulse transmission \cite{jiang_zigzag_2020}, and simplified asymmetric approaches employing calibrated intensity ratios to compensate for channel loss imbalances in practical topologies \cite{wang_simple_2020}. Multi-mode coherent phase-coded TF-QKD has been investigated for spectral efficiency and compatibility with existing optical infrastructure, leveraging standard telecommunication components for cost-effective integration \cite{chistiakov_feasibility_2019}. Hybrid schemes combining TF-QKD with entangled coherent states (ECS) have also been explored to extend entanglement reach and fidelity across long distances \cite{zhang_improving_2019}.

From a deployment standpoint, coherent-state-based TF-QKD has emerged as a cost-effective alternative to single-photon implementations \cite{yin_coherent-state-based_2019}, while asymmetric designs with independently optimized parameters have maintained near-symmetric performance under realistic conditions \cite{grasselli_asymmetric_2019}. Finally, advances in phase stabilization, adaptive intensity modulation, and streamlined sifting collectively enhance secure key rates, system stability, and efficiency, validating TF-QKD as a scalable and high-performance candidate for next-generation quantum communication networks \cite{zhang_improving_2019}.

\begin{table*}[t]
\caption{Comprehensive overview of TF-QKD variants. Security proof status: A = asymptotic, FK = finite-key, C = composable. Distances are based on reported demonstrations, not theoretical limits. Abbreviations: SNS = Send-or-Not-Send; AOPP = All Odd-Parity Pairing; PM-QKD = Phase-Matching Quantum Key Distribution}
\label{tab:comprehensive_tfqkd}
\centering
\begin{tabular}{|p{2.9cm}|p{2.5cm}|p{4cm}|p{2cm}|p{2cm}|p{2cm}|}
\hline
\textbf{Implementation Family} & \textbf{Protocol Variant} & \textbf{Key Innovation} & \textbf{Decoy States} & \textbf{Stabilization} & \textbf{Security Proof} \\
\hline 
\textbf{Traditional TF-QKD} 
& Post-Selection TF-QKD & Original TF-QKD approach with phase-slice post-selection \cite{lucamarini_overcoming_2018, shan_sending-or-not-sending_2024} & Yes (weak+vacuum) & Phase slices ($M$) & A/FK  \\
\hline
\multirow{4}{*}{\textbf{SNS-QKD Family}} 
& SNS-QKD & Send vs. Not-Send decoy strategy optimizing detection efficiency \cite{wang_beating_2019} & Yes  & Tight frequency lock & A/FK/C \\
\cline{2-6}
& Asymmetric SNS-QKD & Channel-adaptive protocols for unequal-loss scenarios \cite{zhou_asymmetric_2019, he_asymmetric_2020, grasselli_asymmetric_2019} & Yes  & Tight frequency lock & A/FK \\
\cline{2-6}
& AOPP-Enhanced SNS-QKD & Odd-Parity Pairing mechanism for enhanced error rejection \cite{zhou_sending-or-not-sending_2024}& Yes  & Tight frequency lock & A/FK  \\
\cline{2-6}
& Passive Decoy-State SNS-QKD  & Hardware-simplified decoy generation using passive beam-splitters \cite{liu_efficient_2022, xue_sending-or-not-sending_2022} & Yes (passive)  & Tight frequency lock & A/FK  \\
\hline
\multirow{3}{*}{\textbf{Phase-Matching Family}} 
& PM-QKD & Coherent phase-matching eliminating post-selection overhead \cite{ma_phase-matching_2018} & Yes  & Tight frequency lock & A/FK  \\
\cline{2-6}
& No-Post-Selection TF-QKD & Complete elimination of phase-slice sifting procedures \cite{cui_twin-field_2019} & Yes & No slicing; phase lock & A/FK  \\
\cline{2-6}
& Discrete-Phase-Randomized TF-QKD & Finite-set phase randomization for simplified control \cite{lorenzo_twin-field_2021} & Yes & Discrete randomization & A/FK  \\
\hline
\end{tabular}
\end{table*}


In order to balance phase control, complexity, asymmetry tolerance, and key performance while maintaining the distinctive $\sqrt{\eta}$ scaling and composable security, TF-QKD has developed into a modular framework that includes PM-QKD, SNS-QKD, and hybrid variants. The PM-QKD provides simplicity with precise phase control, SNS-QKD guarantees strong long-distance performance without phase post-selection but with greater complexity, and hybrid variants improve stability and practicality at the expense of additional overhead. These advances prepared the ground for a more thorough analysis of TF-QKD security, where the fundamental ideas guaranteeing composable security. 

\section{TF-QKD Security}\label{TF:security_proofs}

Security proofs for TF-QKD form the theoretical foundation that guarantees its capability to surpass the linear key rate bound over long distances. Early analyses often worked under asymptotic assumptions, where infinite key lengths, perfectly stable lasers, and symmetric channels allowed mapping prepare-and-measure TF-QKD to equivalent entanglement-based protocols, enabling the application of established tools such as the loss-tolerant framework and phase error rate bounding from observed statistics \cite{krawec_new_2024, curty_simple_2019}. Phase matching was interpreted as an effective basis choice, with the phase error $e_p$ derived from coherent-state overlaps or decoy-state photon-number yields. Simplified proof approaches have also been developed, requiring only a minimal set of observed parameters, making them adaptable to variants like SNS-QKD, discrete-phase protocols, and asymmetric links without major re-derivations \cite{krawec_new_2024, curty_simple_2019}.

Recognizing that realistic systems operate with finite data, modern proofs adopt composable security frameworks valid against general coherent attacks, explicitly accounting for finite-size effects \cite{jiang_unconditional_2019, jiang_composable_2021, huang_effective_2025}. These frameworks employ statistical tools such as multiplicative Chernoff bounds and random sampling without replacement to estimate single-photon yields and phase error rates under limited data. For SNS-QKD, for example, the security failure probability $\epsilon$ is tracked across parameter estimation, error correction, and privacy amplification, with results showing that secure key generation beyond 500 km is possible with realistic pulse numbers ($10^{11}$-$10^{12}$), although optimal decoy intensities differ from asymptotic values due to statistical trade-offs \cite{jiang_unconditional_2019, huang_effective_2025}.

A major trend is the incorporation of implementation imperfections directly into security models. This includes phase reference instability, which reduces interference visibility and influences phase error estimation; intensity fluctuations, which bias decoy-state analysis if unmodeled; asymmetric channel loss, which requires separate treatment of yields and phase errors for each arm; and correlated errors arising from environmental drift or systematic misalignment, which violate i.i.d.\ assumptions and demand correlation-robust bounds \cite{grasselli_asymmetric_2019, lu_practical_2019, panchumarthi_detecting_2023}. Proofs that integrate live monitoring data and fluctuation-aware decoy-state estimation can maintain security under worst-case realizations of these imperfections \cite{grasselli_asymmetric_2019, qiao_sending-or-not-sending_2020}.

Security analyses have also been expanded to account for device-specific and protocol-tailored attacks that go beyond generic adversary models. For example, wavelength-switching attacks exploit imperfect filtering in OPLL-based systems to extract phase reference information, requiring adapted proofs with explicit leakage models to adjust phase error bounds and decoy-state estimates \cite{peng_practical_2025}. Optimized beam-splitting strategies, which combine quantum memory storage with adaptive measurements tailored to the phase-matching structure of TF-QKD, and intensity-mismatch exploitation, where Eve leverages differences between nominal and actual pulse intensities, similarly require refined, attack-aware proofs \cite{wang_optimized_2020, peng_practical_2025}. Countermeasures such as narrow-band filtering, random wavelength switching, and monitoring detectors become more effective when incorporated directly into the proof assumptions rather than treated purely as engineering fixes.

Protocol-specific proof developments reflect the diversity of TF-QKD variants. In SNS-QKD, composable finite-key proofs have been developed with complete statistical fluctuation treatment for ultra-long-distance operation \cite{jiang_unconditional_2019, huang_effective_2025}; discrete-phase protocols achieve security without phase post-selection by directly bounding $e_p$ from discrete phase-randomized decoy states \cite{xu_discrete-phase-randomized_2021}; source-monitoring approaches using Hong-Ou-Mandel interference or light source monitoring help constrain untrusted source behavior \cite{sun_source_2023, qiao_sending-or-not-sending_2020}; and asymmetric link configurations require joint treatment of statistical and intensity fluctuations to avoid overestimating secure rates \cite{grasselli_asymmetric_2019}.

Across these works, the philosophy of TF-QKD security proofs has evolved from idealized, structure-blind analysis toward implementation-aware, attack-specific, and composable models. Although simplified proofs lower the barrier to adaptation and understanding \cite{krawec_new_2024, curty_simple_2019}, practical deployment requires proofs tightly coupled to experimental realities that account for noise sources, finite data, asymmetry, and side-channel leakage. This interplay between theoretical rigor and engineering constraints is central to ensuring that TF-QKD's advantage over the point-to-point repeaterless bound is not just theoretical but robustly secure in practice.

\section{TF-QKD Experimental Realization}\label{TF_experiment}

Since the initial theoretical proposal of TF-QKD in 2018, a substantial number of experimental implementations have demonstrated its potential to surpass the linear rate-loss bound without quantum repeaters. The first proof-of-principle demonstrations \cite{zhong_proof--principle_2019}, experimental TF-QKD has progressed along three main fronts, i.e., to generate truly "twin" fields over long distances, to keep them phase-aligned through lossy and noisy fibers, and to package all of this into deployable systems including field trials, chip-scale integration, and robust security engineering. Early laboratory implementations established that independent lasers can be made to interfere stably at a central, untrusted relay and that TF-QKD’s key rate scaling can beat the repeater-less linear bound. One landmark proof-of-principle system employed a master-slave laser scheme with optical phase-locked loops, time-multiplexed bright reference and weak quantum parts, and fast feedback phase modulators. It achieved interference visibility close to 99.8\% on the bench and 97\% after 300 km of fiber, surpassing the linear bound with a per-pulse key rate of about $6.46\times 10^{-6}$ \cite{wang_beating_2019}. Closely related foundational work validated TF-type operation with decoy states, while the first experimental realization of SNS-QKD confirmed in the lab up to about 300 km that this encoding also beats the linear bound and is robust against source fluctuations \cite{liu_experimental_2019}.

A central challenge is stabilizing the phase between independent lasers and long fibers. One approach relies on a master laser or frequency dissemination and then corrects fiber-induced phase drift. Signal-processing studies quantified drift rates up to 14.5 rad/ms over 1,689 km and showed how strong reference pulses plus FPGA feedback can stabilize interference beyond 800 km with over 96\% visibility for more than two days, eliminating the need for optical amplifiers that would add noise; this yielded 368.9 bps at 511 km and stable operation to 833.8 km \cite{chen_twin-field_2021}. A plug-and-play design moved the laser to the untrusted relay and sent strong pulses out and back through Faraday mirrors, automatically compensating frequency, polarization, and timing with nearly 99\% visibility over 50 km, though residual counter-propagating drift remained limiting at longer spans \cite{park_research_2020, xue_plug-and-play_2021}. Other work replaced shared lasers entirely, the open-channel stabilization multiplexed a reference wavelength with the quantum signal in the same fiber and locked the differential phase at the relay without service fibers, enabling SNS-QKD with 146.7 bps at 403 km, 14.38 bps at 518 km, and 0.32 bps at 615 km, all above the linear bound, while handling even 100 km link asymmetry \cite{zhou_open_2023}. Independent optical-frequency-comb sources, without any common seed, enabled a 546 km field test with several bps and multiday stability \cite{zhou_independent-optical-frequency-comb-powered_2024}. Local frequency reference methods phase-locked each user’s local reference to its own quantum laser, sustaining greater than 97\% visibility with independent lasers and avoiding frequency dissemination \cite{chen_twin-field_2024}. Other demonstrations used purely digital signal processing to estimate and track phase and frequency entirely from multiplexed references \cite{wang_signal_2022}, and subsequent experiments eliminated global phase tracking \cite{zhou_experimental_2023} or explicit phase locking altogether by post-compensating drift in software while still beating the bound \cite{li_twin-field_2023}. These studies collectively mapped the trade-offs between reference-pulse strength, multi-path interference, and estimator performance, showing that careful optimization for example 115 photons per pilot pulse at 834 km keeps pilot noise comparable to dark counts while maintaining high visibility \cite{chen_twin-field_2021}.

Progress soon extended from spools to real field networks. The first SNS-QKD field test over 428 km demonstrated stable operation under real environmental noise \cite{liu_field_2021}, and a city-to-city 511 km deployment showed secure key exchange in carrier fiber \cite{chen_twin-field_2021}. Distance records advanced further using ultralow-loss fiber and superconducting nanowire single-photon detectors: operation at 830 km \cite{wang_twin-field_2022}, nearly 1000 km \cite{liu_experimental_2023}, and finally 1002 km with full finite-key analysis, marking a critical step toward practical continental-scale security \cite{liu_1002_2023}. Another deployment combined TF-QKD with distributed acoustic sensing on the same 658 km link, showing coexistence with classical sensing functions and active mitigation of vibration noise \cite{chen_quantum_2022}. For a better understanding of real networks, which are often asymmetric, with one arm shorter or cleaner than the other, experiments have addressed imbalance. Using a Sagnac loop and the CAL19 asymmetric-intensity strategy, it was shown that intentionally choosing unequal signal intensities outperforms adding artificial loss to the short arm. At 50 dB total loss with a 10 dB imbalance, this method delivered key rates on the order of $10^{-5}$ per pulse, while other methods failed, and even at 56 dB only the asymmetric-intensity method retained positive key rates. Also, reinforced that TF-QKD is robust for provided intensities and feedback are tuned to the imbalance \cite{zhong_proof--principle_2021}, and open-channel stabilization likewise sustained tens of bps under strong asymmetry \cite{zhou_open_2023}.

In order to move toward compact, practical devices, demonstrations have been made on photonic chips. In one SNS-QKD system, a master laser at the relay optically injection-locked on-chip slave lasers and uses active stabilization, achieving high-visibility interference and secure key rates surpassing the linear bound in the asymptotic regime. A complementary experiment implemented phase encoding and injection locking directly on InP chips, achieving about 99\% visibility and validating compatibility with wavelength division multiplexing (WDM) and chip-scale preparation \cite{du_twin-field_2024}. These results indicate a path to low-cost, compact TF-QKD transceivers with integrated sources and modulators. Finally, security engineering considerations have been tested experimentally. Hong-Ou-Mandel interference has been used for continuous source monitoring, diagnosis of indistinguishability, and catching fluctuations or tampering without degrading performance \cite{sun_source_2023}. Another study identified correlated errors, such as detector bias drifts or channel noise, that can bias finite-key analysis if ignored; incorporating correlation checks restores secure margins in real experiments \cite{panchumarthi_detecting_2023}. Together with SNS protocol’s inherent resilience \cite{liu_experimental_2019, liu_field_2021}, these methods strengthen TF-QKD against practical side channels.

Collectively, these experimental advances confirm that TF-QKD is no longer confined to laboratory feasibility studies, but is rapidly approaching operational maturity. With demonstrated robustness under asymmetric losses, environmental perturbations, and real-world deployed fibers alongside integration into photonic chips and plug-and-play configurations, TF-QKD stands as one of the most promising architectures for scalable, long-distance, and cost-effective quantum key distribution.

 \section{TF-QKD Implementation and Deployment} \label{TF:Implement}
As TF-QKD transitions from laboratory proof-of-principle to real-world applications, several practical factors must be addressed. In this section, we outline the key hardware requirements, network architectures, and classical integration strategies, and discuss the experimental progress in realizing practical TF-QKD testbeds.
 

The practical realization of TF-QKD over long distances imposes stringent requirements on both hardware and network architecture. At the physical layer, ultra-stable narrow-linewidth lasers (on the order of a few kHz) are essential to maintain phase coherence between distant transmitters. These are paired with active phase-locking systems, low-drift frequency references, and stabilization loops capable of mitigating environmental perturbations. High-efficiency superconducting nanowire single-photon detectors (SNSPDs) with low dark count rates and fast recovery times are critical to achieving high key rates while minimizing QBER. Furthermore, low-loss optical fiber, polarization-maintaining components, and temperature-controlled spools are deployed to suppress channel fluctuations, with integrated photonic chips increasingly being used to reduce the system footprint and enhance stability.

In terms of network architecture, initial TF-QKD deployments are likely to adopt point-to-point configurations linking two trusted nodes, typically over dedicated dark fibers or WDM channels coexisting with classical traffic. For extended coverage, chains of trusted-node TF-QKD links can bridge metropolitan to intercity scales while preserving realistic security assumptions. As the technology matures, multi-user quantum networks may adopt mesh or star topologies anchored at central measurement nodes, enabling flexible endpoint pairing, optimized resource allocation, and efficient key distribution to diverse users.

Integration with classical infrastructure is vital for large-scale rollout. Precision timing systems, such as White Rabbit or Precision Time Protocol (PTP) provide synchronization for phase references and detector gating across nodes, with classical synchronization signals often co-propagated alongside quantum channels. Generated quantum keys must feed seamlessly into classical Key Management Systems (KMS) to enable automated refresh, authentication, and secure delivery to higher-layer applications, including virtual private networks (VPNs) and transport layer security (TLS) accelerators. Software-defined networking (SDN) frameworks can unify classical and quantum control planes, allowing centralized performance monitoring, fault diagnosis, and real-time reconfiguration in response to channel degradation or environmental changes.

The roadmap to large-scale TF-QKD deployment is expected to follow a staged evolution. Pilot trials over 10-50 km metro-area testbeds will validate stability under environmental variations and coexistence with classical data. Within 3-5 years, regional backbones spanning 100-200 km could emerge, interlinking multiple cities through chained TF-QKD segments and trusted nodes integrated into national research and education networks. 5-10 year horizon envisions national and even international meshes spanning telecom backbone fibers, with automated inter-operator key management and compliance to quantum security standards such as ETSI QKD and ITU-T recommendations. Beyond a decade, commercial-scale deployment is projected to offer rack-mounted, turnkey TF-QKD modules delivering on-demand secure channels for critical infrastructure, financial institutions, and governmental communications.

Achieving stable operation at scale requires careful tuning of multiple practical levers. These include selecting an optimal laser linewidth and frequency-locking method whether offset locking or optical frequency comb referencing designing phase-tracking loops with bandwidths in the kHz-MHz range depending on fiber length and turbulence, and implementing pilot-tone insertion with appropriate wavelength offsets, modulation formats, and guard bands. Additional considerations involve maintaining high extinction ratios for both intensity modulators and decoy states, as well as ensuring interferometer stability through stringent thermal drift budgets and active stabilization where necessary. By harmonizing robust hardware platforms, adaptive network architectures, and integrated classical-quantum management systems, TF-QKD can be transformed from an experimental breakthrough into a cornerstone of a future quantum-secure internet.

Real-world deployment requires a careful balance between adaptive network design, precise hardware, and smooth classical integration, as demonstrated by the practical implementation of TF-QKD. Practical issues such as low-noise detection, synchronized timing systems, and stable phase coherence serve as the cornerstones of dependable operation, and scalable architectures from national backbones to metro testbeds that showcase its increasing maturity. This indicates that TF-QKD is ready to move from experimental setups to a key element of international quantum-secure communication networks with further developments in photonic integration, synchronization, and key management.

\section{TF-QKD in Quantum Networks}\label{TF-Networks}

Extending TF-QKD from point-to-point links to quantum networks and multiparty scenarios has emerged as a major research focus, driven by the need for scalable, long-distance, and secure quantum communication infrastructures. Recent developments span protocol innovations, hardware architectures, and network-level optimization, all aimed at leveraging TF-QKD’s hallmark $\sqrt{\eta}$ distance scaling in more complex topologies.

Advances in protocol development of TF-QKD have directly led to multi-user scenarios and multi-party security proofs. 
Early works on hybrid QKD architectures integrate TF-QKD with conventional decoy-state QKD, enabling star-topology networks where users can flexibly switch between high-rate short-distance links and long-distance TF-QKD without trusted relays \cite{begimbayeva_hybrid_2024, begimbayeva_development_2024}.
A hybrid QKD framework integrates TF-QKD as the core quantum communication primitive within multi-user network topologies, combining DV and CV techniques to maximize compatibility and performance \cite{begimbayeva_development_2024}. To support equitable participation, fair multi-party TF-QKD protocols introduce scheduling and sifting methods that guarantee uniform key rates across users with differing channel losses, ensuring network fairness without sacrificing throughput \cite{abhignan_twin-field-based_2025}. 
TF-QKD principles have also been extended to quantum conference key agreement (QCKA) \cite{carrara_overcoming_2023}, where all participants share a common group key. Protocols such as TF-QKD-inspired CKA surpass the fundamental repeater-less bounds for multi-party settings, and coherent one-way QCKA schemes combine COW encoding with TF-QKD interference to achieve efficient, one-way multi-party key generation \cite{zhong_simple_2022}. Extensions of TF-QKD to QCKA enable a group of \(N\) parties to share a common key with security against general attacks, by using phase-randomized coherent states interfered at an untrusted station and processed with generalized decoy-state analysis, the protocol surpasses the linear rate-loss limit for QCKA, achieving \(O(\sqrt{\eta})\) scaling in total channel transmittance. Simulations indicate secure key generation for up to six users over 300 km even with finite keys, and flexible reconfiguration without full network re-initialization makes it attractive for dynamic multiparty scenarios. Similarly, single-photon interference CKA protocols adapt TF-QKD’s central interference model for scalable group conferencing, offering composable security against general attacks \cite{grasselli_conference_2019}. This dual-layer approach mitigates insider threats and enables differentiated security levels for broadcast versus private communications. Network simulations indicate that incorporating TF-QKD into WDM channels within metropolitan optical backbones can sustain high aggregate key rates under realistic finite-key constraints, while enabling flexible routing and dynamic key allocation under SDN control. These scalability oriented designs tackle the challenge of serving many users simultaneously. The use of  constraint-free high-rate TF-QKD protocols removes the need for probabilistic user scheduling and strict intensity calibration, enabling all users to transmit in parallel with decoy-state compensation for mismatched intensities \cite{xie_scalable_2023}. 

Further enhancements for the stability and practicality of hardware. Sagnac-interferometer-based architectures have provided a phase-stable method for TF-QKD networks. This is exploited in bidirectional transmission in a single fiber loop to provide passive phase stabilization across multiple users, significantly reducing environmental sensitivity and active feedback requirements \cite{mandil_long-fiber_2025}. These architectures enable robust long-distance interference without dedicated phase-reference channels, simplifying deployment in field environments. 
A shared unbalanced Sagnac loop among multiple access nodes enables any user pair to implement TF-QKD without active phase stabilization between distant lasers.  Integration with WDM and time-slot allocation supports concurrent multi-user operation. Experiments over a 407.3 km fiber loop demonstrate secure key generation consistent with \(O(\sqrt{\eta})\) scaling, and simulations suggest feasibility beyond 600 km under realistic detector efficiencies. The 2$\times $N TF-QKD configurations allow dense user packing and dual-hub redundancy, making metropolitan-scale quantum networks feasible with minimal additional fiber infrastructure \cite{park_2n_2022}. At the other end of the complexity spectrum, simple multi-user TF-QKD designs employ a shared light source and passive phase stability to minimize per-user hardware, lowering deployment barriers for early-stage or cost-sensitive networks \cite{zhong_simple_2022}. 

Building on the features of hardware, the network-level optimization addresses the operational layer of multi-user TF-QKD. Switching strategies, both static and dynamic, have been investigated to maximize aggregate key rates, maintain fairness, and minimize idle time when users share access to a central measurement station. Optimization frameworks using mixed-integer programming and heuristics show that adaptive switching policies can significantly improve throughput and latency over fixed schedules, making them essential for large-scale TF-QKD backbone networks \cite{karavias_switching_2024}. Collectively, these advances illustrate a multi-pronged evolution of TF-QKD from a two-party protocol into a foundation for quantum-secure network infrastructures.

\section{Open Problems and Future Directions}
\label{open_problems}
The maturation of TF-QKD opens several promising avenues for realizing scalable and resilient quantum networks. These research directions span technological development, architectural integration, and deployment frameworks, each of which is critical for advancing TF-QKD from experimental feasibility to a cornerstone of global quantum communication.

\subsection{Satellite and Free-Space Integration}
A major frontier for TF-QKD lies in extending its reach to free-space and satellite channels, thereby enabling global coverage. Such integration requires overcoming atmospheric turbulence, beam wandering, and variable transmittance through adaptive protocols and real-time channel assessment. Key technical challenges include designing low-Earth orbit (LEO) quantum satellites with onboard measurement modules, compensating for Doppler-induced phase fluctuations, and constructing hybrid satellite--terrestrial networks. Experimental demonstrations of TF-QKD over realistic satellite--ground links remain a primary milestone toward practical deployment.

\subsection{Quantum Memory and Hybrid Architectures}
Scaling TF-QKD beyond the fundamental $O(\sqrt{\eta})$ limit necessitates integration with quantum memories and repeater-based architectures. Quantum memories enable temporal storage and synchronization, thereby improving throughput and error correction capabilities. Furthermore, hybrid designs that combine TF-QKD backbone links with entanglement-swapping repeaters present opportunities for continental-scale communication. Achieving this vision requires the development of long-coherence, high-fidelity quantum memories, as well as purification protocols that are compatible with TF-QKD infrastructure.

\subsection{Automation and Intelligent Optimization}
Operational stability remains a critical challenge for real-world TF-QKD systems. Machine learning--based automation offers a powerful solution, with reinforcement learning approaches providing adaptive phase and polarization stabilization beyond traditional feedback methods. Predictive algorithms may also enable proactive compensation for environmental disturbances. On the security side, anomaly detection systems can identify implementation vulnerabilities, while adaptive protocol optimization supports real-time threat response. Such intelligent control will reduce operational overhead and enhance the scalability of TF-QKD networks.

\subsection{Scalable Network Architectures}
Transitioning from point-to-point links to large-scale networks requires innovations in scalability, flexibility, and resilience. Software-Defined Quantum Networking (SDQN) offers a control framework for multi-user environments, while quantum-aware routing protocols and efficient WDM are essential for coexistence with classical traffic. In addition, fault-tolerant topologies are necessary to ensure secure connectivity despite individual link or node failures, paving the way toward robust and flexible quantum network infrastructures.

\subsection{Standardization and Deployment Frameworks}
The path to commercial adoption depends heavily on standardization and certification. Efforts through organizations such as ETSI and ITU-T are vital to establish interoperability across TF-QKD variants, define composable finite-key security guarantees, and create performance benchmarks that allow fair comparisons across implementations. International harmonization of these standards will facilitate seamless interconnection of emerging infrastructures, ensuring the foundation for a globally consistent quantum-secure internet.

Collectively, these research directions highlight the transition of TF-QKD from laboratory-scale demonstrations to commercial-grade communication infrastructure. Progress in satellite integration, hybrid memory-assisted architectures, intelligent automation, scalable network design, and standardization will ultimately determine TF-QKD’s role as a foundation technology for future quantum-secured networks.

\section{Conclusion}\label{conclusion}

This comprehensive survey has examined the transformative impact of TF-QKD on long-distance quantum communication, documenting its evolution from theoretical breakthrough to practical implementation over the past seven years. Our analysis reveals that TF-QKD fundamentally alters the rate-distance relationship in quantum communication by achieving $R \sim \sqrt{\eta}$ scaling, effectively circumventing the PLOB bound that limited all previous repeaterless protocols to linear scaling. This represents a qualitative shift that extends secure communication distances from hundreds to potentially thousands of kilometers, positioning TF-QKD as the leading candidate for next-generation quantum networks.

The development of robust variants, including PM-QKD, SNS-QKD, and hybrid implementations, demonstrates remarkable adaptability to diverse deployment scenarios. Each variant addresses specific implementation challenges while preserving the fundamental scaling advantage, with SNS-QKD emerging as particularly robust against environmental fluctuations and hardware imperfections. The transition from asymptotic to finite-key security proofs represents a critical milestone in practical quantum cryptography, where modern composable security frameworks explicitly account for statistical fluctuations, implementation imperfections, and side-channel vulnerabilities, ensuring that theoretical advantages translate to real-world security guarantees.

Record-breaking experimental demonstrations exceeding 1000 km in fiber, coupled with field trials in metropolitan networks, confirm that TF-QKD has transitioned from laboratory curiosity to deployment-ready technology. The integration of advanced stabilization techniques, photonic chip implementations, and compatibility with existing telecommunications infrastructure validates the protocol's commercial viability. TF-QKD's significance extends beyond incremental performance improvements to enable fundamentally new network architectures. By supporting intercity and potentially intercontinental quantum links without intermediate trusted nodes, TF-QKD provides the backbone infrastructure necessary for a global quantum internet. The protocol's MDI security model addresses critical trust concerns in distributed networks, while its compatibility with classical telecommunications infrastructure facilitates practical deployment.

Despite remarkable progress, several research frontiers remain critical for widespread TF-QKD deployment. Seamless integration with SDN frameworks, automated key management systems, and existing network operations centers requires continued development. Long-term stability under diverse environmental conditions remains a key challenge, where advanced stabilization techniques leveraging machine learning for predictive compensation and autonomous system reconfiguration represent promising research directions. The development of cost-effective implementations through photonic integration, simplified architectures, and shared infrastructure utilization will determine the pace of commercial adoption, while the maturation of international standards through ETSI, ITU-T, and other bodies will be crucial for ecosystem development.

As we look toward the next decade, TF-QKD stands at the threshold of transforming secure communication from specialized research applications to ubiquitous infrastructure. The protocol's unique combination of enhanced security, extended reach, and practical implementability positions it to serve as the quantum analog of classical fiber-optic networks that revolutionized telecommunications in the late 20th century. The convergence of maturing TF-QKD technology with emerging quantum applications, including distributed quantum computing, quantum-enhanced sensing networks, and advanced cryptographic protocols, suggests a future where quantum advantage extends far beyond key distribution.

TF-QKD represents more than an incremental advance in quantum communication, it embodies a paradigm shift that makes practical quantum-secured networks achievable with current technology. By overcoming fundamental physical limitations while maintaining rigorous security guarantees, TF-QKD provides a clear pathway from laboratory demonstrations to global deployment. The rapid pace of theoretical development, experimental validation, and practical implementation documented in this survey demonstrates the exceptional momentum behind TF-QKD research and development. As governments, enterprises, and research institutions increasingly recognize quantum technologies as strategic priorities, TF-QKD's proven performance and deployment readiness position it as the foundation for the emerging quantum communication infrastructure. The future quantum internet will require robust, scalable, and secure communication protocols capable of supporting diverse applications across global distances, through its breakthrough scaling properties and mature security framework, provides the technological foundation necessary to realize this vision, with continued advancement playing a pivotal role in determining humanity's readiness for the quantum age of secure communication.


\bibliographystyle{ieeetr}
\bibliography{refs}
\end{document}